%% file: main.tex
\documentclass[oneside,11pt]{article}
\input{macros}

\title{\vspace{-1.5cm}The latent cognitive structures of social networks}

\author{Izabel Aguiar\\ Stanford University \\ \texttt{izzya@stanford.edu} 
      \and
      Johan Ugander\\ Stanford University\\ \texttt{jugander@stanford.edu}}
\date{}
\begin{document} 
\maketitle

\begin{abstract}
\input{abstract}
\end{abstract}
\section{Introduction} \label{sec:intro}
\input{intro}

\section{Related Work}\label{sec:sc}
\input{related_work}
\section{A factor model of cognitive social structures}\label{sec:ntd}
\input{nntuck}
\subsection{Statistical tests for cognitive structure}\label{sec:tests}
\input{tests}
\input{limitations}
\section{Empirical cognitive structure}\label{sec:krack} 
\input{application}
\section{Conclusion}\label{sec:conclusion}
\input{conclusion}

\section*{Funding}
IA acknowledges support from the NSF GRFP and the Knight-Hennessy Scholars Fellowship. JU acknowledges partial support from ARO (\#76582-NS-MUR) and NSF (\#2143176).

\section*{Declaration of competing interests}
The authors declare that they have no known competing interests that could have influenced the work in this paper.

\section*{Acknowledgements}
We sincerely thank the two anonymous reviewers who provided generous and thoughtful feedback to make this manuscript stronger and more clear. We thank Keith Hunter for making his data available and for his generous availability for discussions. We thank Amir Goldberg and our MURI collaborators for their fruitful discussions and encouragement on this work. 

\section*{Data Availability} The friendship \acl{css} from \citet{krackhardt1987} and the employees' attribute data is available in the \verb|statnet| R package \citep{handcock2003}. A transcription of the advice \acl{css} from \citet{krackhardt1987} is available on the Github repository of the present work, as well as that of \citep{aguiar2022factor}. The advice and friendship longitudinal \acs{css} datasets are generously available online \citep{hunter2019}. Tools for this work, a notebook reproducing all figures and table values in this manuscript, as well as a tutorial notebook (all in python) can be found at \verb|https://github.com/izabelaguiar/CSS|.

\clearpage

\bibliography{mybib}
\bibliographystyle{plainnat}

\input{appendix}
\clearpage

\end{document}

%% file: macros.tex
\usepackage[headheight=1in,margin=1.1in]{geometry}
\usepackage{lipsum}
\usepackage{siunitx}
\usepackage{multirow}
\usepackage{amsfonts}
\usepackage{graphicx}
\usepackage{epstopdf}
\usepackage{appendix}
\usepackage{natbib}
\usepackage{amssymb}
\usepackage{fancyhdr}
\usepackage{algorithmic}
\usepackage{hyperref}
\hypersetup{hidelinks}
\usepackage[symbol]{footmisc}

\usepackage{bm}
\setcitestyle{round}
\newtheorem{definition}{Definition}
\usepackage[acronym]{glossaries}
\input{glossaries}
\usepackage[capitalize,nameinlink]{cleveref}[0.19]
\usepackage{xcolor}
\usepackage{algorithm}[0.1]

\newcommand{\ten}[1]{\boldsymbol{\mathcal{#1}}}
\newcommand{\mat}[1]{\bm{#1}}
\renewcommand{\vec}[1]{\mathbf{#1}}

\newcommand{\R}{\mathbb{R}}
\newcommand{\G}{\boldsymbol{\mathcal{G}}}
\newcommand{\M}{\boldsymbol{\mathcal{M}}}
\newcommand{\A}{\boldsymbol{\mathcal{A}}}

\newcommand{\U}{\bm{U}}
\newcommand{\V}{\bm{V}}
\newcommand{\Y}{\bm{Y}}
\newcommand{\I}{\bm{I}}

\newcommand{\Adims}{N \times N \times L}
\newcommand{\Gdims}{K \times K \times C}

\newcommand{\Aspace}{\mathbb{Z}_{0+}}
\newcommand{\Gspace}{\mathbb{R}_{+}}
\newcommand{\by}{\times}
\newcommand{\uf}[1]{_{(#1)}}
\newcommand{\Ahat}{\boldsymbol{\mathcal{\hat{A}}}}

\newcommand{\acs}[1]{\acrshort{#1}}
\newcommand{\acf}[1]{\acrfull{#1}}
\newcommand{\acl}[1]{\acrlong{#1}}

%% file: glossaries.tex
\glsdisablehyper
\newacronym{ntd}{NNTuck}{nonnegative Tucker decomposition}
\newacronym{ntds}{NNTucks}{nonnegative Tucker decomposition}
\newacronym{ntdcap}{NNTuck}{Nonnegative Tucker Decomposition}
\newacronym{sbm}{SBM}{stochastic block model}
\newacronym{sbms}{SBMs}{stochastic block models}
\newacronym{lrt}{LRT}{likelihood ratio test}
\newacronym{lrts}{LRTs}{likelihood ratio tests}
\newacronym{css}{CSS}{cognitive social structure}
\newacronym{csspl}{CSSs}{cognitive social structures}
\newacronym{cssplc}{CSSs}{Cognitive social structures}
\newacronym{mt}{MT}{MULTITENSOR}
\newacronym{slrt}{split-LRT}{split likelihood ratio test}
\newacronym{nmf}{NMF}{nonnegative matrix factorization}
\newacronym{xval}{cross-validation}{cross-validation}
\newacronym{em}{EM}{expectation maximization}
\newacronym{dcmmsbm}{dc-mm-SBM}{degree-corrected, mixed-membership SBM}
\newacronym{pmf}{PMF}{Poisson matrix factorization}
\newacronym{kld}{KL-divergence}{Kullback-Leibler divergence}

\newacronym{capntd}{NNTuck}{Nonnegative Tucker Decomposition}

\newacronym{capsbm}{SBM}{Stochastic Block Model}
\newacronym{capsbms}{SBMs}{Stochastic Block Models}

\newacronym{capnmf}{NMF}{Nonnegative Matrix Factorization}

\newacronym{capem}{EM}{Expectation Maximization}

\newacronym{cappmf}{PMF}{Poisson Matrix Factorization}

\newacronym{sca}{SCA}{social-cognitive agreement}

%% file: abstract.tex
When people are asked to recall their social networks, theoretical and empirical work tells us that they rely on shortcuts, or heuristics. Cognitive Social Structures (CSS) are multilayer social networks where each layer corresponds to an individual's perception of the network. 
With multiple perceptions of the same network, CSSs contain rich information about how these heuristics manifest, motivating the question, \textit{Can we identify people who share the same heuristics?} In this work, we propose a method for identifying \textit{cognitive structure} across multiple network perceptions, analogous to how community detection aims to identify \textit{social structure} in a network. To simultaneously model the joint latent social and cognitive structure, we study CSSs as three-dimensional tensors, employing low-rank nonnegative Tucker decompositions (NNTuck) to approximate the CSS\textemdash a procedure closely related to estimating a multilayer stochastic block model (SBM) from such data. We propose the resulting latent cognitive space as an operationalization of the sociological theory of \textit{social cognition} by identifying individuals who share \textit{relational schema}. In addition to modeling cognitively \textit{independent}, \textit{dependent}, and \textit{redundant} networks, we propose a specific model instance and related statistical test for testing when there is \textit{social-cognitive agreement} in a network: when the social and cognitive structures are equivalent. We use our approach to analyze four different CSSs and give insights into the latent cognitive structures of those networks.

%% file: intro.tex
The study of social networks often concerns itself with the notion of a \textit{true} network. For a given relationship type, the true network is typically defined by what is captured when asking each individual about their own relationships \citep{wasserman1994social}. A common extension of this approach is to uncover multiple true networks by using multiple name generators \cite[e.g.,][]{campbell1991, banerjee2013}. Yet other approaches use interaction data to contrast physical contact networks to participants' faulty recollections \citep[e.g.,][]{killworth1976informant, bernard1979informant, freeman1987cognitive}. The study of \acf{csspl} \citep{krackhardt1987} (sometimes referred to simply as ``Krackhardt data''), where \textit{perceptions} of the social network are collected from each individual within it, makes a radical departure from these approaches. Krackhardt proposed the idea that individuals' perceptions provide valuable information about the network, and through them, multiple notions of the network exist. Much of the subsequent body of work on \acs{csspl} has since focused on measuring how these perceptions differ from some sense of a common or underlying network. In this work we stray from the conception of a true network, and instead intend to learn what perceived networks tell us about the joint social and cognitive structures that \acs{csspl} capture and what we can learn from considering these varied perceptions as a whole.

\acl{cssplc} capture information present, but usually ignored, in our social networks: our (varied) perceptions of the social structure. \citet{newcomb1961acquaintance} first introduced the concept of incorporating this perspective in social network analysis by asking participants about ties involving both themselves and others, and \citet{krackhardt1987} expanded this perspective in his work that laid the foundations for the formal study of \acl{csspl}. For the purposes of this work we will consider the \acs{css} as a \textit{multilayer network}, wherein each layer is defined by a different individual's perception of the social network \citep[see][for a review of multilayer networks]{de2013mathematical}.

The present work makes the following contributions. First, our work proposes examining the \acs{css} of a population to study the latent structure of the perceptions of individuals within the network. We approach this task using nonnegative Tucker decompositions. The \acf{ntd}, which decomposes a tensor into three nonnegative \textit{factor matrices} and one nonnegative \textit{core tensor}, is a multilayer network extension of the \acl{dcmmsbm} proposed in \citet{BallKarrerNewman2011}. Analogous to how the factor matrices in the single layer \acs{sbm} identify {\it node communities}, the additional third factor matrix in the \acs{ntd} identifies {\it layer communities}. As such, the third factor matrix of the \acs{ntd} allows for the adjacency tensor to be low rank in the layer dimension, and identifies interdependencies between the layers of a multilayer network. In the case of the \acs{css} multilayer network, the \acs{ntd} identifies interdependencies between the \textit{cognitions} of the network.

Second, we introduce the \acs{ntd} as a useful operationalization of the sociological theories of \textit{relational schema} \citep{baldwin1992} and \textit{social cognition} \citep{howard1994social}, theories that each discuss the ways in which we carry and perceive social structures in our minds, and the shortcuts we use to do so. We propose that the \acs{ntd} can be used as a tool for identifying sets of individuals with commonly shared relational schema, doing so in an unsupervised manner without deciding a priori \textit{which} schema to examine \citep[as do, e.g.,][]{janicik2005social, kilduff2008organizational}. The interpretation of the relational schema at play can instead be interpreted after the latent cognitive spaces have been identified. 

Third, we propose statistical tests for assessing different types of cognitive structure in a social network. Notably, we propose a definition and test for determining if the latent cognitive structure is mirrored by (or mirrors) the social structure, a property we call {\it \acl{sca}}. Finally, and most broadly, our work examines the \acs{css} as a three dimensional tensor, connecting the study of \acs{csspl} to useful formalisms from multilinear algebra for studying such rich data objects. We use the tools developed here to uncover insights into the cognitive and social spaces of empirical networks, showing how the \acs{ntd} can be used to explore how social and cognitive structure can significantly alter across time.

The structure of the work is as follows. In \Cref{sec:sc} we discuss related work. We introduce the \acf{ntd} of multilayer networks \citep{aguiar2022factor} in \Cref{sec:ntd}, where we also discuss the sociological concepts that we draw upon and propose the interpretation of the \acs{ntd} in the context of analyzing \acs{csspl}. In \Cref{subsec:SCA} we define and discuss the notion of \textit{\acl{sca}} useful for analysing \acs{css} datasets, and \Cref{sec:tests} we present for statistical tests for assessing latent cognitive structure in social networks. In \Cref{sec:krack} we investigate the use of the \acs{ntd} to find latent structure in four \acs{css} datasets from \citet{krackhardt1987} and \citet{hunter2019}. We conclude with a discussion and promising open questions for future work. 

%% file: related_work.tex
This work is by no means the first which suggests that there is a latent, or shared, cognitive structure present in the perceptions of social networks. Even preceding Krackhardt's original \acs{css} work, 
\citet{carley1986approach} showed how the shared ``frames'' developed in the social networks of the members of a college dormitory impacted shared cognition.  In \citet{krackhardt1987}, the author motivates his proposal of \acl{csspl} by discussing schemas, invoking research that finds people rely on patterns for their recollections \citep{freeman1987cognitive}, and suggesting that \citeauthor{heider1958}'s (\citeyear{heider1958}) balance theory is a possible cognitive model underlying peoples' perceptions. Later, \citeauthor{dimaggio1997culture}'s (\citeyear{dimaggio1997culture}) work on \textit{Culture and Cognition} cites \citeauthor{krackhardt1987}'s (\citeyear{krackhardt1987}) work, stating that ``networks are crucial environments for the activation of schemata, logics, and frames.''

Although there was a strong emphasis on the theory of shared schema leading up to and within \citet{krackhardt1987}, the bulk of the \acs{css} work that has followed has focused on quantifying the ``accuracy'' of the perceptions of members in a social network. For example, \citet{krackhardt1990assessing} found that those with more accurate perceptions of the network are perceived as being more powerful, and \citet{brands2014just} relate the misperceptions of women's roles in a network to the characteristics that their peers attribute to them. For a comprehensive review of \acs{css} work through \citeyear{brands2013cognitive}, see \citet{brands2013cognitive}.

Our work connects four separate literatures: (i)~modeling latent structure in a \acs{css}, (ii)~studying relational schema in network perceptions, (iii)~multilayer network extensions of the \acl{sbm}, and (iv)~identifying layer interdependence in a multilayer network. As such, we discuss related work in the literature of each of these four categories.

\paragraph{Latent space models of CSSs.} From a modeling perspective, our work is similar to work which has aimed to identify latent structure in the perceptions of the network by considering the \acs{css} data object in its entirety. In \citet{kumbasar1994systematic}, the authors conduct a correspondence analysis of a $N^2 \times N$ ``stacked'' matrix representation of a \acs{css} to find and compare the latent self-perceived and group-perceived social spaces of each person. In \citet{sosa2021latent}, the authors propose a 3-dimensional extension of \citeauthor{hoff2002}'s (\citeyear{hoff2002}) latent space model for social networks. Their model identifies the self-perceived latent position of each individual, and the posterior probabilities of a Bayesian model are used to assess ``whether the perception of an individual about its own position in social space agrees with the judgments of other actors.'' The authors propose that this information can be used to create a weighted (single layer) network, where the weights of an edge between persons $i$ and $j$ corresponds to how much cognitive agreement $i$ and $j$ share. 
Relatedly, in \citet{sewell2019latent} the author proposes a latent space model which describes the deviations between individuals' perceptions through bias and variance, explaining individuals' tendencies to report more or less dense networks, and their confidence in the network perception, respectively. 
More recently, \citet{debacco2023} proposed a latent model to uncover an unobserved underlying network from multiply reported network data (such as a \acs{css}) by identifying latent propensities for each reporter to over or under report edges between others, as well as each reporter's tendency to report reciprocated ties between people. These related papers which have focused on identifying latent spaces in a \acs{css} set the precedent that we hope to continue here. In contrast to this previous work, however, the present work centers a latent space model which also describes a generative process of the varying perceptions, in order to identify who in the network shares relational schema. 

\paragraph{Relational schema in network perceptions.} From a conceptual perspective, our work is most similar to that of \citet{janicik2005social}, \citet{carnabuci2018}, \citet{menon2014}, \citet{kilduff2008organizational}, \citet{emery2011relational}, \citet{brashears2013humans}, and \citet{brashears2015}, which all aim to identify specific relational schema individuals use when recalling their social networks. In contrast to our proposed method, these prior works all identify \textit{specific} relational schema to study the presence of in the social network, whereas we aim to identify \textit{people} who share a general relational schema that can be interpreted and identified \textit{a posteriori}. \citet{janicik2005social} specify two relational schema that influence social network perceptions: balance schema \citep{heider1958} and linear-ordered schema \citep{desoto1959}. Their work goes on to study how people use these schema to fill in incomplete parts of a social network. \citet{carnabuci2018} take an experimental approach to infer the relational schema that participants rely upon by measuring how learning rates change in conditions which assume different relational schema in the network. Similarly, \citet{menon2014} study how psychological priming impacts the density and characteristics of people's perceived networks. \citet{kilduff2008organizational} study multiple \acs{css} datasets to understand how schemas like small-world assumptions \citep{milgram1967small} and perceived popularity manifest themselves in \acs{csspl}. In \citet{emery2011relational}, the goal is to identify how specific relational schemas impact the emergence of leadership in a social network. In \citet{brashears2013humans}, the author conducts experiments to understand how compression heuristics aid in the recall of large social networks, finding that people have better recall when structural assumptions of relational schema like kinship and triadic closure are accurate. In \citet{brashears2015} the same experimental data is reanalyzed and the recall of triads and groups is compared to what is expected from exponential random graphs. We aim to enrich the study of these varied questions about cognitive structure. In contrast to this previous work, the operationalization we introduce identifies shared relational schema without necessitating \textit{which} relational schema to identify ahead of time, and enables us to contrast social structure with cognitive structure.

\paragraph{Multilayer extensions of the stochastic block model.}The model we propose for analyzing \acs{csspl} builds upon a diverse set of work on \acf{sbms} for multilayer networks. Most similarly to the \acs{ntd} work of \citet{aguiar2022factor} is that of \citet{schein2016bayesian} and \citet{deBacco}. In \citet{deBacco}, a multilayer \acs{sbm} is proposed by estimating a separate \acs{sbm} for each layer of the network, keeping the social space fixed across layers. In \citet{schein2016bayesian}, the authors propose a Bayesian Poisson Tucker Decomposition (BPTD) as a generalization of the \acf{dcmmsbm} \citep{BallKarrerNewman2011} to study multilayer networks. The model proposed in \citet{deBacco} is a specific instance of the \acs{ntd} wherein the third factor matrix in the decomposition is constrained to be the identity matrix (see \Cref{def:ind} in \Cref{sec:ntd} for more details). Whereas \citet{schein2016bayesian} estimate the \acs{sbm} with an MCMC algorithmic approach, the \acs{ntd} multiplicative updates procedure seeks to maximize the log-likelihood to obtain a point estimate of the same model (see \Cref{subsec:ntd} and \Cref{nntuck:alg} for more details). See \citet{aguiar2022factor} for more discussion on how the \acs{ntd} relates to and generalizes other multilayer \acs{sbms}.

\paragraph{Layer interdependence in multilayer networks.}Viewing the \acs{css} more generally, as a multilayer network, identifying structure in the cognitive space (which we propose to do with the \acs{ntd}) is similar to identifying \textit{layer interdependence}, for which there have been a multitude of proposed methods. Beginning in the \acs{css} literature, \citet{krackhardt1987} suggested differentiating layer similarity by comparing individual layers to a \textit{consensus structure}. \citet{battiston2014structural} and \citet{kao2018layer} introduce and use various similarity measures to identify layer communities. \citet{de2015structural} and \citet{de2016spectral} cluster similar layers using information-theoretic tools. In \citet{stanley2016}, layers are categorized into groups that were drawn from the same \acs{sbm}. In \citet{deBacco} the authors build multiple models using different subsets of the layers and layer interdependence is determined via each models' performance on a link prediction task. In contrast to this previous work, the \acs{ntd} identifies layer interdependence by identifying \textit{layer communities} which have the same underlying generative model. As such, the \acs{ntd} allows for the adjacency tensor to be low rank in the layers as well as the nodes. Because identifying layer interdependence in a \acs{css} with the \acs{ntd} relates the varying perceptions to one another via a shared generative process (as opposed to this other work which uses similarity measures or information-theoretic tools), it is thus a natural operationalization of existing sociological theories.

%% file: nntuck.tex
In this section we propose the use of the \acf{ntd}  \citep{aguiar2022factor} for analyzing \acl{csspl}. We introduce the \acs{ntd} as a multilayer \acf{sbm}, discuss the interpretation of the \acs{ntd} in the context of \acs{csspl}, and relate it to relevant sociological theories.

\subsection{The nonnegative Tucker decomposition} \label{subsec:ntd}
The \acf{dcmmsbm} \citep{BallKarrerNewman2011} is a generative model of a single layer network that assumes each of the $N$ nodes in the network have soft (``mixed'') membership in each of $K$ different communities, and that an affinity matrix captures the Poisson rates at which nodes in each community form an edge with one another. More specifically, for a network of $N$ nodes represented by an adjacency matrix $A \in \Aspace^{N \times N}$, the \acs{dcmmsbm} assumes that each node $i$ has a nonnegative  outgoing membership vector, $\vec{u}_i \in \R_+^K$, and a nonnegative incoming membership vector, $\vec{v}_i \in \R_+^K$. Each membership vector represents the node's soft assignment to $K\leq N$ groups when forming outgoing and incoming edges, respectively. The $K\by K$ nonnegative affinity matrix $\mat{G}$ describes the rates of connections between nodes in each communitiy. Then, for $N \by K$ matrices $\U$ and $\V$ with rows defined by $\vec{u}_i$ and $\vec{v}_i$, the \acs{dcmmsbm} assumes that $\mat{A}\sim \text{Poisson}(\U \mat{G} \V^\top)$. While the model allows for adjacency matrices $\mat{A}$ consisting of nonnegative count data, it is common practice to employ the \acs{dcmmsbm} for model-based analyses of binary network data. Note that constraining all the elements of $\vec{u}_i,  \vec{v}_i,$ and $\vec{G}$ to be nonnegative allows for these parameters to be interpretable as membership weights and affinities. 

The \acf{ntd} as a generative model is a multilayer network extension of the \acs{dcmmsbm}. It builds on related work on multilayer \acs{sbms} \citep[e.g.,][]{schein2016bayesian, deBacco, tarres2019tensorial} by generalizing the \acs{sbm} to a multilayer setting and using tensor decompositions to jointly model the layers, much like how matrix decompositions underlie traditional (single layer) stochastic block models. 

Consider a multilayer network with $N$ nodes and $L$ layers represented by adjacency tensor $\A \in \Aspace^{\Adims}$. The \acs{ntd} multilayer \acs{sbm} again assumes that each node $i$ has nonnegative membership vectors $\vec{u}_i \in \R_+^K$ and $\vec{v}_i \in \R_+^K$, representing the node's soft assignment to $K\leq N$ groups when forming outgoing and incoming edges, respectively (note that for an undirected network, $\vec{u}_i = \vec{v}_i$). Furthermore, the \acs{ntd} assumes that each layer $\ell$ has a nonnegative vector $\vec{y}_\ell \in \R_+^C$ describing the \textit{layer's} soft membership to each of $C\leq L$ \textit{layer communities}. Just as matrices $\U$ and $\V$ in the \acs{dcmmsbm} describe latent community structure in the \textit{nodes} of single-layer networks, the factor matrix $\Y$ in the \acs{ntd} describes latent structure in the \textit{layers} of a multilayer network. Finally, the \acs{ntd} assumes a nonnegative core tensor $\G \in \Gspace^{\Gdims}$ whose frontal slices are $C$ different affinity matrices. 
Let $\vec{u}_i, \vec{v}_i, \vec{y}_\ell$ be the rows of nonnegative matrices $\U,\V,$ and $\Y$, respectively. Then the \acs{ntd} multilayer \acs{sbm} assumes that
\begin{equation}
    \A \sim \text{Poisson}(\G \times_1 \U \times_2 \V \times_3 \Y).
    \label{Tucker}
\end{equation}
When estimating this model from multilayer network data, maximizing the log-likelihood of observing $\A$ under the model given by \cref{Tucker} is equivalent to minimizing the \acs{kld} between $\A$ and $\Ahat = \G \times_1 \U \times_2 \V \times_3 \Y$. This equivalence is essential to why the \acs{ntd} estimated by minimizing the \acs{kld} is synonymous with the multilayer \acs{sbm}.

\subsection{Interpretation of the NNTuck of a CSS}\label{subsec:ntd_css}
Through its multiple factor matrices, the \acs{ntd} multilayer \acs{sbm} models both the latent social structure\textemdash sometimes referred to as the Blau space \citep{blau1977macrosociological,mcpherson1983ecology,mcpherson1991evolution}\textemdash and the latent layer structure of a multilayer network. When viewing a \acs{css} as a multilayer network, each layer corresponds to a different individual's perception of the social network. Therefore, the latent layer structure captured by the \acs{ntd} of a \acs{css} is precisely the structure identified amongst the varied perceptions of the network, what we hereafter refer to as the \textit{latent cognitive structure} of the \acs{css}. In the setting of \acs{csspl}, then, the \acs{ntd} identifies three latent spaces in a network: the outgoing and incoming social groups in the $\U$ and $\V$ factor matrices, respectively, and the $\Y$ factor matrix, which identifies $C$ \textit{cognitive groups} and each individual's membership to each of these groups. 

\begin{figure}
    \centering
    \includegraphics[width = .7\textwidth]{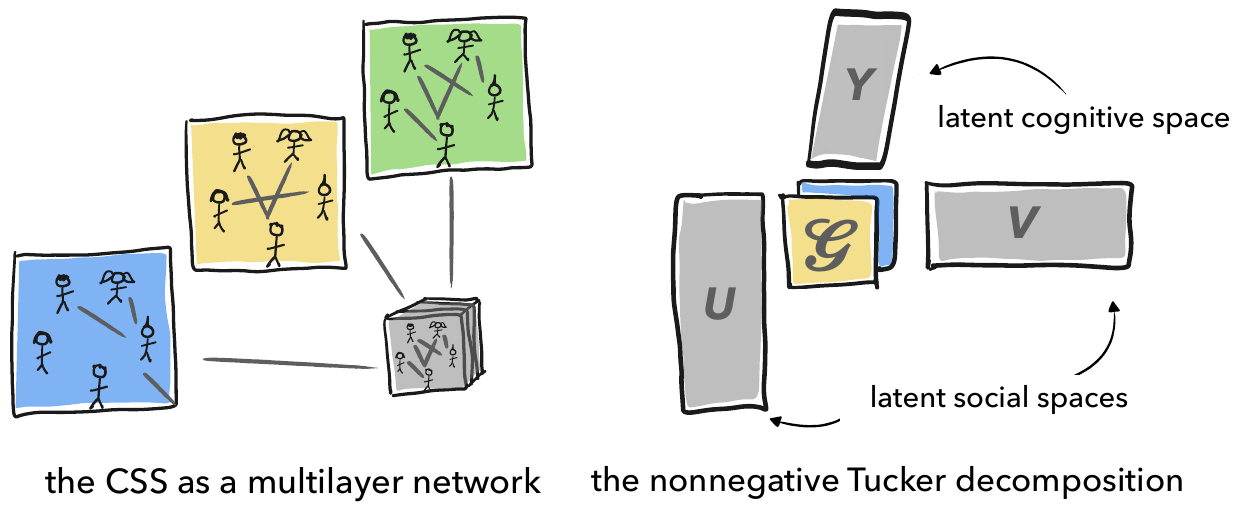}
    \caption{In this work we analyze \acf{csspl} as multilayer networks represented by an $N \by N \by N$ adjacency tensor (left). The \textit{frontal slices} of the adjacency tensor are visualized in blue, yellow, and green. The $N$ frontal slices of the \acs{css} are adjacency matrices representing the perception each person has of their network. We use the \acf{ntd} to model the \acs{css} with a multilayer \acl{sbm}, which decomposes the adjacency tensor into latent social spaces and a latent cognitive space (right).}
    \label{fig:tensor_intro}
\end{figure}

\paragraph{Relationship to Social Cognition and Relational Schema.} 
Aiming to interpret this factor model through sociological theories, we will not review all relevant or adjacent sociological literature surrounding cognition, social representations, schema, or culture's various impact on the three. While tempting, it is far beyond the scope of the present work. Instead, we focus on connecting the \acs{ntd} to \citeauthor{baldwin1992}'s (\citeyear{baldwin1992}) theory of \textit{relational schema} and \citeauthor{howard1994social}'s (\citeyear{howard1994social}) theory of \textit{social cognition}.

A succinct summary of \citeauthor{baldwin1992}'s theory of \textit{relational schema} (emphasis our own) is that,
\begin{quote}
\textit{``[relational schema] describe \textbf{expectations about the nature of relationships between people}, defining what aspects of social interactions individuals will pay attention to, as well as \textbf{the attributes of others that are meaningful} within those interactions''}\citep{brands2013cognitive}.
\end{quote}
Relational schema describe the generalizations or patterns we rely on when making assumptions about relationships or when making new social connections. In data collection for \acl{csspl}, each person is asked to report on their perceptions of the relationships between every pair of people in their network. It has been empirically shown that people rely on compression heuristics when recalling large networks \citep{brashears2013humans}, and it is reasonable to assume that the perceptions reported in \acs{csspl} are the product of some such compression heuristics. In \Cref{subsec:hunter}, we use the \acs{ntd} to understand the cognitive latent space of a longitudinal \acs{css} collected repeatedly over several weeks, allowing us to investigate the heuristics involved (and not involved) in that early formation of the cognitive social structure for that population.

About \textit{social cognition}, \citeauthor{howard1994social} writes,
\begin{quote}
\textit{    ``Social cognition articulates explicitly how social structures are carried in individuals' mental systems \ldots The content of the group dimensions on which people are prone to categorize reveals the connections between cognitive and social structures.''}
\end{quote}
\acs{css} datasets provide empirical access to ``how social structures are carried in individuals' mental systems,'' and we propose that the \acs{ntd}, through the $\Y$ factor matrix, can identify specific ways in which any differences in these perceptions can be attributed to differences in such mental systems. 

\begin{figure}
    \centering
    \includegraphics[width = .95\textwidth]{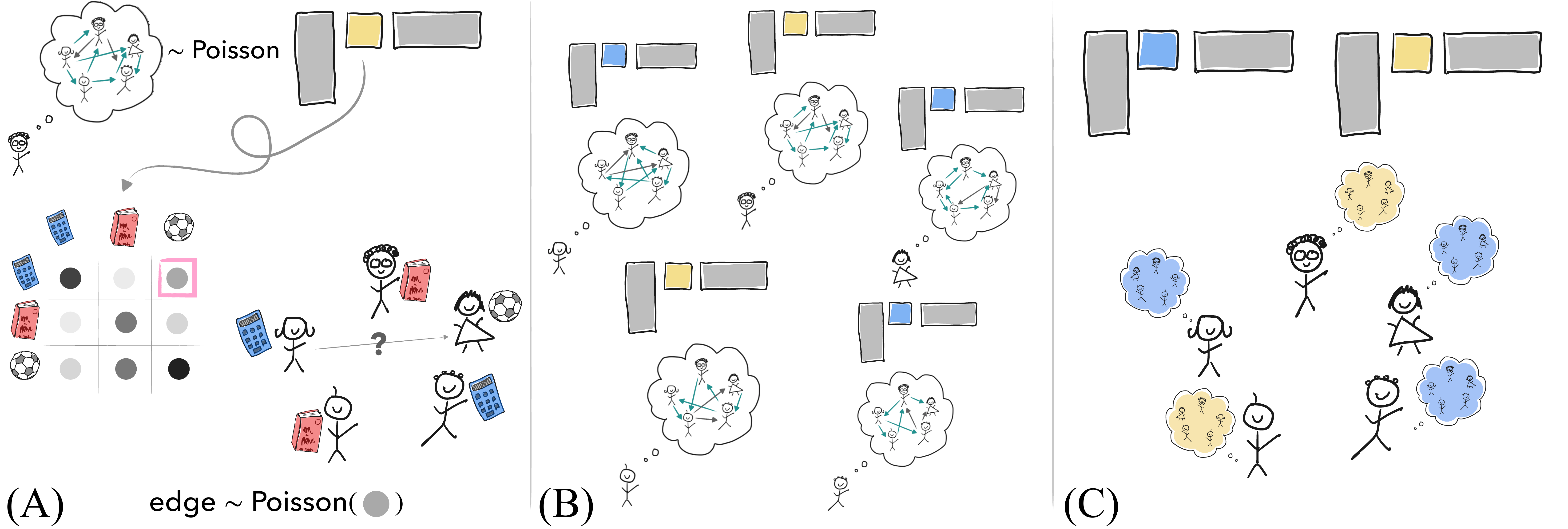}
    \caption{A visualization and example showing the connection between the \acf{sbm}, the \acf{ntd}, and relational schema. (A) We assume that each person generates their perception of the network according to a stochastic block model (SBM). A person’s perception of the existence of an edge is drawn according to the rate specified by the affinity matrix in each person’s SBM. Previous empirical and theoretical work on how people store and recall large social networks suggests that a coarse model of relationship, like the SBM, well represents this cognitive process. (B) We \textit{could} estimate a separate affinity matrix to describe each person’s network perception. (C) However, the NNTuck allows us to identify when people share the same generative process for their perceptions, interpretable as sharing the same relational schema.}
    \label{fig:css_schematic}
\end{figure}

Because the \acs{ntd} is an extension of the \acs{sbm} to multilayer networks, it serves as a \textit{generative model} of multilayer networks, where the role of the $\Y$ factor matrix is to identify layers which can be modeled using the same generative \acs{sbm}. In the context of the \acs{css}, the $\Y$ factor matrix identifies individuals in the network who rely on the same generative process when recalling their perception of the network. As such, we propose that individuals in the same cognitive group can be thought of as having a shared set of relational schema. In this sense, too, the \acs{ntd} provides an explicit operationalization of \citeauthor{howard1994social}'s theory of social cognition, because it jointly models the connection between the cognitive and social structure in the network.  In \Cref{sec:krack} we interpret possible relational schema underlying the shared cognitive spaces in three empirical datasets, and in \Cref{subsec:hunter}, we directly tie our analysis using the \acs{ntd} to \citeauthor{howard1994social}'s theories. For a visualization of the connection between the \acs{ntd} and relational schema, see \Cref{fig:css_schematic}.

To formalize our interpretation of the \acs{ntd} in terms of established social theories, we now turn to possible different assumptions of the structure and dimension of the ``layer'' factor matrix $\Y \in \R^{L \by C}$ in the \acs{ntd} of a \acs{css}. The following definitions apply specifications for multilayer network models from \citet{aguiar2022factor} to the \acs{css} context. They form the basis of our primary exploration of empirical \acs{css} data in \Cref{sec:krack}.

\begin{definition}[Cognitively independent \acs{ntd}]\label{def:ind}
A \textbf{cognitively independent \acs{ntd}} is a \acl{ntd} where $C = L$ and $\Y$ has the constraint $\Y = \mat{I}$, meaning each individual in the network relies on a distinct relational schema.
\end{definition}
\begin{definition}[Cognitively dependent \acs{ntd}]\label{def:dep}
A \textbf{cognitively dependent \acs{ntd}} is a \acl{ntd} where $\Y$ has the constraint $C<L$, meaning each individual's social cognition can be described by a mixture of $C$ relational schema.
\end{definition}
\begin{definition}[Cognitively redundant \acs{ntd}] \label{def:red}
A \textbf{cognitively redundant \acs{ntd}} is a \acl{ntd} where $C=1$ and we constrain $\Y$ to be the ones vector, $\Y = [1, \dots, 1]^{\top}$, meaning each individual in the network has the exact same relational schema in perceiving their network.
\end{definition}

\subsection{Social-Cognitive Agreement}\label{subsec:SCA}

In addition to the above model specifications, we next propose a new specification that is particularly well-suited to the CSS context. Considering that the \acs{ntd} of a \acs{css} can tell us both about the latent social and cognitive structures, a natural question to ask is how the two spaces are related to one another. Do our social surroundings directly influence how we conceptualize our social network, does our conceptualization influence how we socialize, and if so, can we empirically assess the ways in which they do? 

The hypothesis that our cognitions are influenced by our peers was proposed in \citet{carley1986approach}, where a main goal of the work was to ``relate cognitive structure to social structure at an empirical level.'' As Carley writes, ``the social and cognitive processes cannot be decoupled.'' Indeed, through a combination of quantitative and qualitative data collection, Carley identifies specific tightly knit groups in a community who share similar cognitive mappings of a concept. A similar idea was later echoed in \citet{freeman1992}, where the author writes, ``the individuals involved in any particular local community would be expected ultimately to produce very similar mental images of group structure in that community.''

These related hypotheses have a natural analogue to a specification of the \acs{ntd}. Namely, the social structure is the same as the cognitive structure if $\U=\Y$. We refer to this structural assumption as one of  \textit{social-cognitive agreement}.
\begin{definition}[Social-cognitive agreement \acs{ntd}] \label{def:sc_agreement}
An \acs{ntd} with \textbf{\acf{sca}} is a \acl{ntd} where $C<L$ and we constrain $\Y=\U$. Individuals in the same social group share the same relational schema.
\end{definition}

We note that constraining $\Y$ and $\U$ to be the same does not imply that everybody in the same social/cognitive group will have the exact same perceptions. Rather, the model is constrained so that the \textit{structure} we observe in the social space is also reflected in the structure of the cognitive space. The implication of this constraint is that the perceptions of people in the same social/cognitive space are modeled with the same generative model, which will result in \textit{similar} (but not the same) perceived networks.

\subsection{Model optimization details} \label{subsec:opt_det}
Algorithmically, our primary method for maximizing the log-likelihood of observing a \acs{css} dataset under a given \acs{ntd} model specification is to employ the multiplicative updates algorithm of \citet{kimChoi}, an effective extension of the multiplicative updates algorithm for \acf{nmf} given by \citet{LeeSeung}. For more details, see \Cref{nntuck:alg} in \Cref{apdx:sca} or \citet{aguiar2022factor}. The \citet{kimChoi} algorithm, like the NMF algorithm it builds on, is guaranteed to find a local maxima of the nonconvex log-likelihood, but not necessarily the global maxima. That said, extensive empirical evaluations have shown it to perform very well when employed using best practices for nonconvex optimization (e.g., choosing the best of many random restarts of the optimization routine).

For the cognitively independent, dependent, and redundant \acs{ntd} model assumptions (\Crefrange{def:ind}{def:red}), this algorithm transfers easily to the relevant specifications. Briefly, when constraining $\Y = \I$ or $\Y = \vec{1}$ as in the cognitively independent and redundant \acs{ntds}, respectively, the algorithm maintains monotonic convergence to a local minimum by simply initializing these constraints and never updating $\Y$ after initialization. Similarly, when constraining $\U = \V$, as in the case of an undirected network, by initializing both $\U=\V$ and symmetry in the frontal slices of $\G$, monotonic convergence to a local minimum is maintained. The symmetric structure in the frontal slices of adjacency tensor $\A$ (as is the case in an undirected network) and core tensor $\G$ (which is an interpretable assumption for an undirected network) result in equivalence between the multiplicative update derived for $\U$ and the multiplicative update derived for $\V$. Thus, by simply initializing a symmetric $\G$ and $\U=\V$, \Cref{nntuck:alg} maintains this symmetry and equivalence.

In the case of estimating a \acl{sca} \acs{ntd} (\Cref{def:sc_agreement}), however, constraining $\Y = \U$ is a nontrivial constraint on the multiplicative updates algorithm. The structure in both the data $\A$ and core tensor $\G$ that would be necessary in order to ensure equivalence in the multiplicative updates of $\Y$ and $\U$ (as we noted for the symmetric network specification above), are contextually unreasonable to expect or assume. Specifically, equivalence in the multiplicative updates for $\Y$ and $\U$ requires that the 1-unfolding and the 3-unfolding\footnote{Tensor unfoldings are higher-order analogues of matrix vectorizations. See \cite{Kolda2009} for explanations of tensor unfoldings and other tensor properties.} of $\A$ are equivalent ($\mat{A}_{(1)}=\mat{A}_{(3)}$) and that the 1-unfolding and the 3-unfolding of $\G$ are equivalent ($\mat{G}_{(1)}=\mat{G}_{(3)}$). In the context of the \acs{css}, this structural assumption on $\A$ means that, for every pair of people $i$ and $j$, person $i$'s perception of $j$'s outgoing relationships is the same as person $j$'s perception of person $i$'s outgoing relationships. In order for this assumption to hold, each person would need to report their \textit{own} outgoing relationships as their perceptions for everyone else's outgoing relationships (that they assume everybody goes to the exact same people for advice/friendship).

Thus, in order to estimate a \acf{sca} \acs{ntd}, seemingly minor changes to the \citet{kimChoi} algorithm nontrivially change its behavior. In considering other algorithmic approaches, \cite{Cambre1999} (and more recently \cite{jin2022}) develop and discuss a method for estimating a \textit{symmetric} Tucker decomposition (with no nonnegativity constraint) wherein all three factors ($\U, \V, \Y$ in our vocabulary), are constrained to be equivalent. However, in this and other related work, the focus is on developing algorithms for decomposing a data tensor which already has an appropriately symmetric structure. Such data symmetry is something that, again, is not reasonable to assume in the case of \acs{csspl}.

As such, we move forward with altering \Cref{nntuck:alg} to accommodate the \acs{sca} constraint, resigning ourselves to a estimation algorithm without guaranteed monotonic convergence to a local optima. That said, this was only ever a weak guarantee relative to guarantees of reaching a global optima, as enjoyed by many optimization procedures (e.g., for convex log-likelihoods). 

We alter the multiplicative updates from \cite{kimChoi} in the following ways. First, we initialize nonnegative factors $\Y, \V$, and nonnegative core tensor $\G$. Then, we set $\U = \Y$. For each multiplicative iteration, we proceed by first updating $\V$, then updating $\U$, and then updating $\Y$, all according to the same multiplicative updates as in \Cref{nntuck:alg}. As our key modification, after updating $\Y$ and before updating $\G$, we set both $\U$ and $\Y$ to be equal to their average. This step ensures that the model remains ``feasible'' within the constrained model class, essentially updating both factor matrices to reflect the gradients in both roles (as $\U$ and $\Y$ factors). We then update $\G$ as usual. As a final minor modification, given the lack of monotonicity guarantee, \Cref{SCAMUNTD} also features a modified termination criteria to ensure the return of the minimum solution found during a given solution trajectory.
We describe this approach in full in \Cref{nntuck:SCAalg} in \Cref{apdx:sca}, and provide a python implementation in our code repository (see our Data Availability Statement). In \Cref{apdx:sca} we also provide further details on our decision to choose this modification over others.

This heuristic optimization procedure for returning a maximum likelihood estimate of an \acs{ntd} with \acl{sca} is provided as an initial feasible approach to the question of modeling \acl{sca} in \acs{csspl}. It is an interesting open question if the multiplicative updates algorithm of \citet{kimChoi}, or some other algorithm, can be modified to produce a maximum likelihood estimate with local or global optimality guarantees. For example, \cite{arora2012computing} showed that under ``approximate separability'' \citep{donoho2003does} assumptions on a data matrix, the \acf{nmf} has a guaranteed solution and an associated convex optimization problem. Their proposed algorithm overcame prior limitations of heuristics for approximating an \acs{nmf} by minimizing the Frobenius norm error of the approximation. Estabilishing identifiability and convergence conditions for the \acl{ntd} has been addressed in, e.g., \cite{xu2015alternating} and \cite{sun2023}. In \cite{xu2015alternating}, the author proposes an algorithm for estimating the \acl{ntd} with global convergence guarantees, but does so for minimizing only the Frobenius loss. In \cite{sun2023}, the authors propose a regularization that makes the \acl{ntd} identifiable, but the associated loss based on \acs{kld} is non-convex. Extending this work to estimating the \acl{ntd} (and the \acl{ntd} with \acl{sca}) with global convergence guarantees under loss given by the \acs{kld} is promising future work. 

%% file: tests.tex
The above structural assumptions are all articulated in terms of constraints on the parameters of the given models, where the constrained models all lie within a subspace of the more general (cognitively independent) model. As such, these assumptions are highly testable from data using likelihood ratio tests. We adapt three statistical tests from \citet{aguiar2022factor} to have vocabulary specific to the interpretation of the \acs{css}, and introduce an additional statistical test for social-cognitive agreement, all for studying the cognitive structure of a social network through its \acs{css}. 

\begin{definition}[Cognitive independence]\label{def:lrt-ind}
For a multilayer network let model I be the cognitively independent \acs{ntd} and let model II be the cognitively dependent \acs{ntd} with $C<L$. A \acs{css} has \textbf{cognitive independence} at level $\alpha$ if the \acl{lrt} (LRT) with $(L-C)K^2 - LC$ degrees of freedom is significant at level $\alpha$.
\end{definition}
\begin{definition}[Cognitive dependence]\label{def:lrt-dep}
A \acs{css} has \textbf{cognitive dependence} at level $\alpha$ if the \acs{lrt} described above is \textit{not} significant at level $\alpha$ for a pre-specified $C$.
\end{definition}
\begin{definition}[Cognitive redundance]\label{def:lrt-red}
A \acs{css} has \textbf{cognitive redundance} at level $\alpha$ if the \acs{lrt} comparing the cognitively redundant \acs{ntd} to the $C=2$ cognitively dependent \acs{ntd} with $K^2 + 2L$ degrees of freedom is not significant at level $\alpha$.
\end{definition}
\begin{definition}[Social-cognitive agreement]\label{def:lrt-sca}
A \acs{css} has \textbf{social-cognitive agreement} at level $\alpha$ if the \acs{lrt} comparing a cognitively dependent \acs{ntd} with $K'$ and $C'$, not necessarily equal, to a \acl{sca} \acs{ntd} with the constraint $K=C$ and $\U = \Y$ with $2N(K'-K)+K'^2C' + C'N - K^3$ degrees of freedom is not significant at level $\alpha$. The cognitively dependent \acs{ntd} must be such that both $C'\geq K$ and $K'\geq K$ is true.
\end{definition}

The use of the standard likelihood ratio test, as in the analogous tests for layer interdependence in \citet{aguiar2022factor}, relies on Wilks' Theorem \citep{wilks1938} and its assumptions. Wilks' Theorem provides the standard theory for the asymptotic validity of likelihood ratio tests.
In the context of statistically comparing different \acs{ntd} models, we do not satisfy the necessary assumptions or regularity conditions in order for Wilks' Theorem to hold: the \acl{ntd} is neither identifiable\footnote{Recent work by \cite{sun2023} has proposed that imposing a ``volume regularization'' condition on the factor matrices of the \acl{ntd} ensures identifiability of the decomposition. Implementing these conditions in contextually suitible ways for \acs{csspl} and, more generally, multilayer networks, is promising future work.} \citep{Kolda2009, chen2020note}, nor can we be certain that our estimation of the \acs{ntd} has reached a global maximum of the log-likelihood. Because the \acs{ntd} is not identifiable, the concept of which models are nested within one another is subtle, as one can re-write the \acs{ntd} by absorbing the parameters of the factor matrices into the core tensor, or vice versa. As such, the notion of nested models and the difference in the number of parameters (and subsequently, determining the degrees of freedom of the $\chi^2$ distribution for the standard \acs{lrt}) is nuanced, especially in the context of a 3-mode tensor decomposition.

As a means of addressing the first shortcoming, we utilize the \acs{slrt} from \citet{wasserman2020}, which does not require identifiability or nested models. Although the \acs{slrt} does not require the same strict regularity conditions as does the standard \acs{lrt}, it does lead to lower powered tests.
Notably, although the the power deficit of the \acs{slrt} is somewhat inherent in the test, it may be exacerbated in our particular setting due to our sample size and the presence of many nuisance parameters in our model\footnote{Although we are concerned with testing how well a particular $\Y$ fits the data, we also have to estimate parameters $\U, \V$, and $\G$ in each model. In this case, $\U, \V$, and $\G$ can be considered nuisance parameters.} \citep{tse2022note}.  In \citet{tse2022note}, \citet{strieder2022choice}, and \citet{spectordiscussion}, the authors propose ways to improve the power when testing statistical models with nuisance parameters. For instance, both \citet{tse2022note} and \citet{strieder2022choice} discuss the importance of an optimal splitting of the data when conducting the \acs{slrt}. In the context of the \acs{css}, wherein we are splitting the data of a 3-mode tensor, we have many choices for how to do so, and our na\"{i}ve approach of splitting the data at random may not be optimal. Determining the optimal split of a data tensor and implementing other suggestions for increasing the power of the \acs{slrt} in the context of the \acs{ntd} is a topic for future work.

To address the second shortcoming, of reaching a global maximum, we use the \acs{ntd} corresponding to the highest log-likelihood over multiple initializations of \Cref{nntuck:alg}. However, when estimating a \acs{sca}~\acs{ntd} we use \Cref{nntuck:SCAalg}, which, as discussed in the previous section, is not monotonic in \acs{kld} nor does it come with any convergence guarantees, even to a local minima. As such, we caution against strong interpretation of the results of the \acs{slrt} when testing for \acl{sca} and again highlight addressing this algorithmic limitation as subject of future work. For more discussion on the standard and split-likelihood ratio tests, see \Cref{apdx:lrts}. 

%% file: limitations.tex
\subsection{Modeling limitations}
We conclude this section by discussing some limitations of using the \acs{ntd} for studying \acs{css} datasets. First, the \acs{ntd} models the \acs{css} using a multilayer \acl{sbm}. As such, the latent cognitive space identifies sets of individuals whose perceptions of the network can be well modeled by the same (degree-corrected, mixed-membership) \acs{sbm}. It is in this sense that we say the \acs{ntd} identifies people who share relational schema, because their perceptions can be modeled by the same generative process. A significant focus of the active literature on \acs{sbms} works to enrich their modeling structure to more realistically model empirical networks, noting that the \acs{dcmmsbm} is still quite coarse \citep[e.g.,][]{peixoto2014hierarchical}. However, as we discuss above, both theoretical social theory as well as experimental and empirical studies on how people store and recall large social networks suggests that a coarse model of relationship well represents this cognitive process \citep[see also, e.g.,][]{mullainathan2008}. In the context of \acs{css} datasets, wherein each person is asked to report their perception of the social network, we believe that the coarse reasoning expressed in the \acs{sbm} models this situation well.

Although the \acs{ntd} does not tell us \textit{what} the relational schema are, we believe that the identified spaces can be contextually interpreted (as we do in the empirical applications below). This modeling approach allows for flexibility in identifying people who share relational schema and interpreting that schema \textit{a posteriori}, as opposed to deciding \textit{a priori} which schema to study in a given dataset, which other work on relational schema necessitates \citep[e.g.,][]{kilduff2008organizational}. However, there may be relational schema that may not be well-described using a \acl{sbm}, and latent space models which support these schema, or other generative models of multilayer network formation, would be better suited for modeling such cases. For example, a relational schema based on the assumption of triadic closure would likely be better modeled by a multilayer extension of the recent \acs{sbm} developments presented in \citet{peixoto2022disentangling}.

With these limitations in mind, we turn to four empirical applications of the \acs{ntd}.

%% file: application.tex
We now use the models and tests introduced in previous sections to analyse four different \acs{css} datasets, with the aim of exploring and interpreting the social and cognitive structures in each. In each of the below datasets, we interpret each person's responses as a layer of a multilayer network, representing the \acs{css} as an $N \times N \times N$ adjacency tensor $\A$, where each layer is represented as a frontal slice of the tensor.

\paragraph{Krackhardt Advice/Friendship} In the original \acs{css} work by \citet{krackhardt1987}, $N=21$ managers in a high tech firm were asked to report on their perceptions of the \textit{advice network} and the \textit{friendship network} of the firm, forming two \acs{css} data sets. We also have information about each employee's affiliation to one of four departments (where the president of the company doesn't belong to any department), their relative hierarchy to one another (president, vice president, supervisor), their tenure at the firm, and their age. In what we call the ``Krackhardt Advice \acs{css}'' each person was asked who approached whom for work-related advice, and in the ``Krackhardt Friendship \acs{css}'' each person was asked who was friends with whom. 

\paragraph{Hunter Friendship} Here, we analyze longitudinal \acs{csspl} from \citet{hunter2019} which tracks the friendship network perceptions of $N=20$ college juniors from around the country in a summer leadership class over the course of six weeks. In each week, each student was asked to report on their perceptions of the friendships between students in the class. We also have data about each student's self-reported gender (in the case of this class, only ``male'' and ``female'' were reported), race, academic major, and undergraduate institution. We also know which of 10 dorm rooms and 4 study groups each student was assigned to. Although a CSS was collected at each of six weeks\footnote{The entire longitudinal CSS for each relationship type from \citet{hunter2019} could be meaningfully analyzed as a fourth order tensor of size $20 \times 20 \times 20 \times 6$, studying temporal factors through tensor decomposition methods. That said, this data constitutes the only presently known instance of such fourth order data, and we consider such an analysis to be beyond the scope of the present work.}, we focus on the \acs{css} data corresponding to the first and sixth week. Going forward, we refer to each dataset as the ``Hunter Friendship, Week One \acs{css}'' and ``Hunter Friendship, Week Six \acs{css}'', respectively.
\\
\\
We analyze each \acs{css} dataset above through a three-step procedure of (1) model selection, (2) statistical testing, and (3) visualization and interpretation.

For (1) model selection, our main goal is to assess which dimensions of the latent spaces ($K$ and $C$) should be used for the analysis. To do so, we set up a \acs{xval} link prediction task. By splitting each dataset into a train and test set, we are able to assess how well each model performs in predicting \textit{unseen} data, which can help account for overfitting that may happen with models that have more parameters. The construction of the \acs{xval} approach is such that for each link prediction task we construct five different \textit{masking tensors} which each split the data into different \textit{train} and \textit{test} sets. For masking tensor $\M$, $\M_{ij\ell} = 1$ indicates that the link between $i$ and $j$ in layer $\ell$ is in the train set. Conversely, $\M_{ij\ell} = 0$ indicates that the link between nodes $i$ and $j$ in layer $\ell$ is in the test set (is \textit{missing} or \textit{unknown}), and will be held out in the estimation of the \acs{ntd}. In the \textit{tubular} link prediction task, which we use here, masking is done tube-wise (in the tensorial sense), meaning edges are always observed or unknown across all layers. Missing link $(i,j)$ in layer $k$ implies that link $(i,j)$ is missing in all layers ($\M_{ijk} = 0 \Rightarrow \M_{ij\ell} = 0, \forall \ell$). For $b$-fold \acs{xval}, tubes $(i,j,\cdot)$ in the tensor are missing with uniform and independent probability $1/b$. We select the \acs{ntd} with the highest training set log-likelihood from 20 runs of the multiplicative updates algorithm with different random initializations. Then, test-AUC is averaged across the five different maskings. This process is repeated for varying model assumptions and dimensions $(K, C)$. Parameter choices for $K$ and $C$ may then be determined by considering the pair with the highest test-AUC or by considering the cost of an increase in test-AUC relative to the complexity of the model (e.g., see our discussion of parameter choice in the Krackhardt Advice \acs{css}).

Next, for step (2), statistical testing, we use the appropriate dimensions $K$ and $C$ determined from the above parameter sweep to perform statistical tests for cognitive redundance, independence, and \acl{sca} using the corresponding \acs{lrts}.

Finally, as step (3), we estimate an \acs{ntd} of the dataset using either \Cref{nntuck:alg} or \Cref{nntuck:SCAalg} (the latter for models that feature \acl{sca}) with the model specification decided by the outcome of the \acs{lrt}. We then visualize and interpret the social and cognitive spaces identified by these models.

\subsection{Krackhardt Advice CSS}
\begin{figure}
    \centering
    \includegraphics[width = 0.5\textwidth]{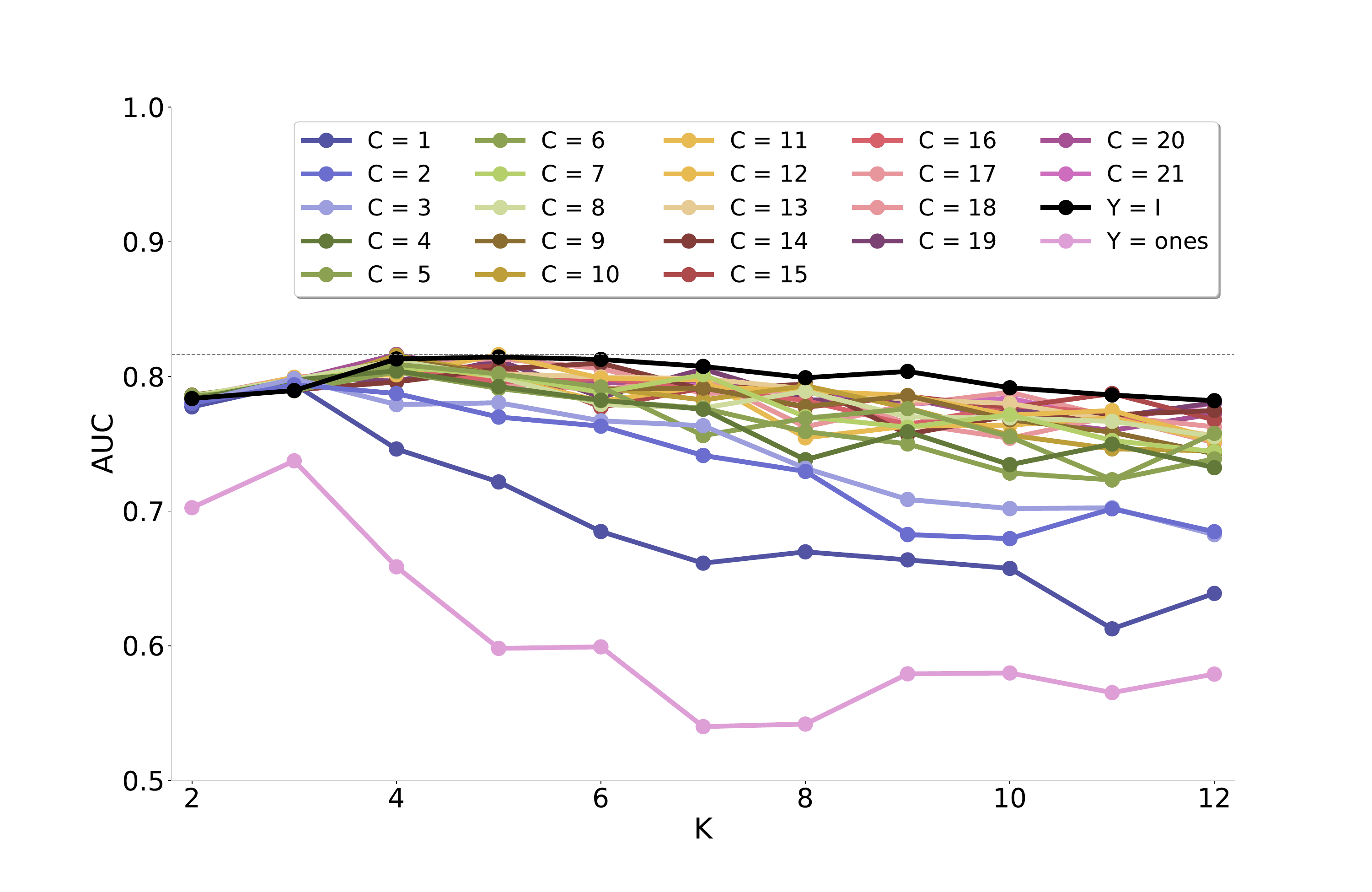}\includegraphics[width = 0.5\textwidth]{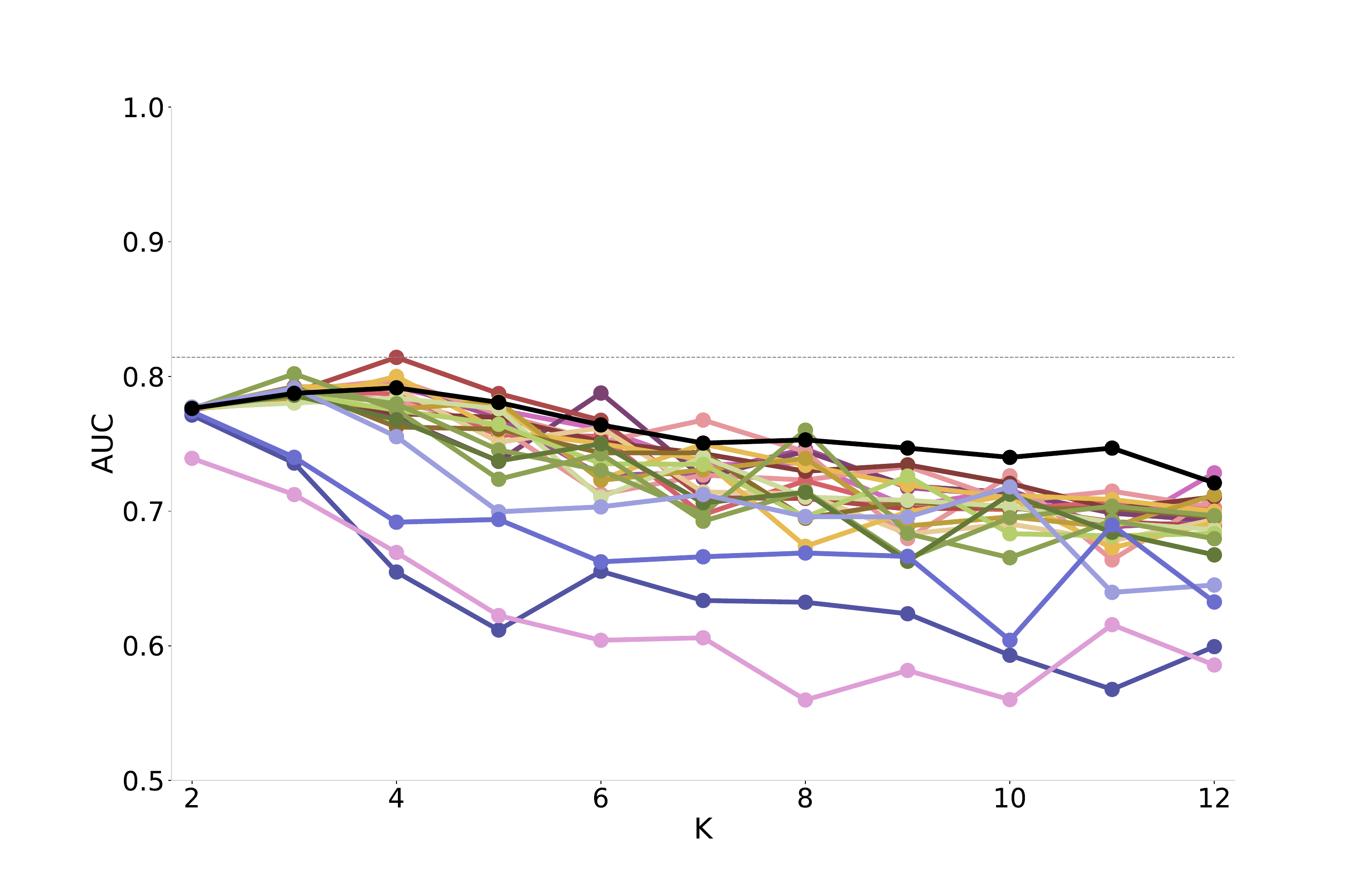}
    \caption{The test-AUC averaged across a tubular fivefold \acs{xval} task for the Krackhardt advice (left) and friendship (right) \acs{css} datasets. The pink and black lines correspond to the \acs{ntd} model assumptions of cognitive redundancy and independence, respectively. Each other colored line corresponds to a different value of $C$ in assuming cognitive dependence in the \acs{css}, and the x-axis corresponds to different choices of the social latent space parameter $K$. Based on this \acs{xval} task, we choose to examine the social and cognitive factor matrices of the advice and friendship \acs{css} datasets corresponding to the cognitively dependent \acs{ntd} with $K=C=3$, and $K = 3,C= 5$, respectively.}
    \label{fig:xval}
\end{figure}

For this \acs{css} with $N=21$ managers, the test-AUC under different model specifications is shown in \Cref{fig:xval} (left). We choose to examine the factor matrices corresponding to the cognitively dependent \acs{ntd} (see \Cref{def:dep}) with $K=3$ and $C=3$. We choose these parameters over others with a higher test-AUC due to the observation that higher values of either $K$ or $C$ result in a minimal increase in test-AUC\footnote{Specifically, considering \Cref{fig:xval} (left), we see that the \acs{ntd} with $K=4$ and $C=20$ has only a slightly higher test-AUC (0.0178 higher) and comes at the cost of an additional 633 parameters.}. We perform a \acl{lrt} to compare the goodness-of-fit of the nested, cognitively dependent \acs{ntd} with $K=C=3$, to the full cognitively independent \acs{ntd}. In doing so, we fail to reject the null hypothesis that the data was generated from the smaller cognitively dependent model (see \Cref{table:LRT}). 

\begin{figure}
    \centering
    \includegraphics[width = .85\textwidth]{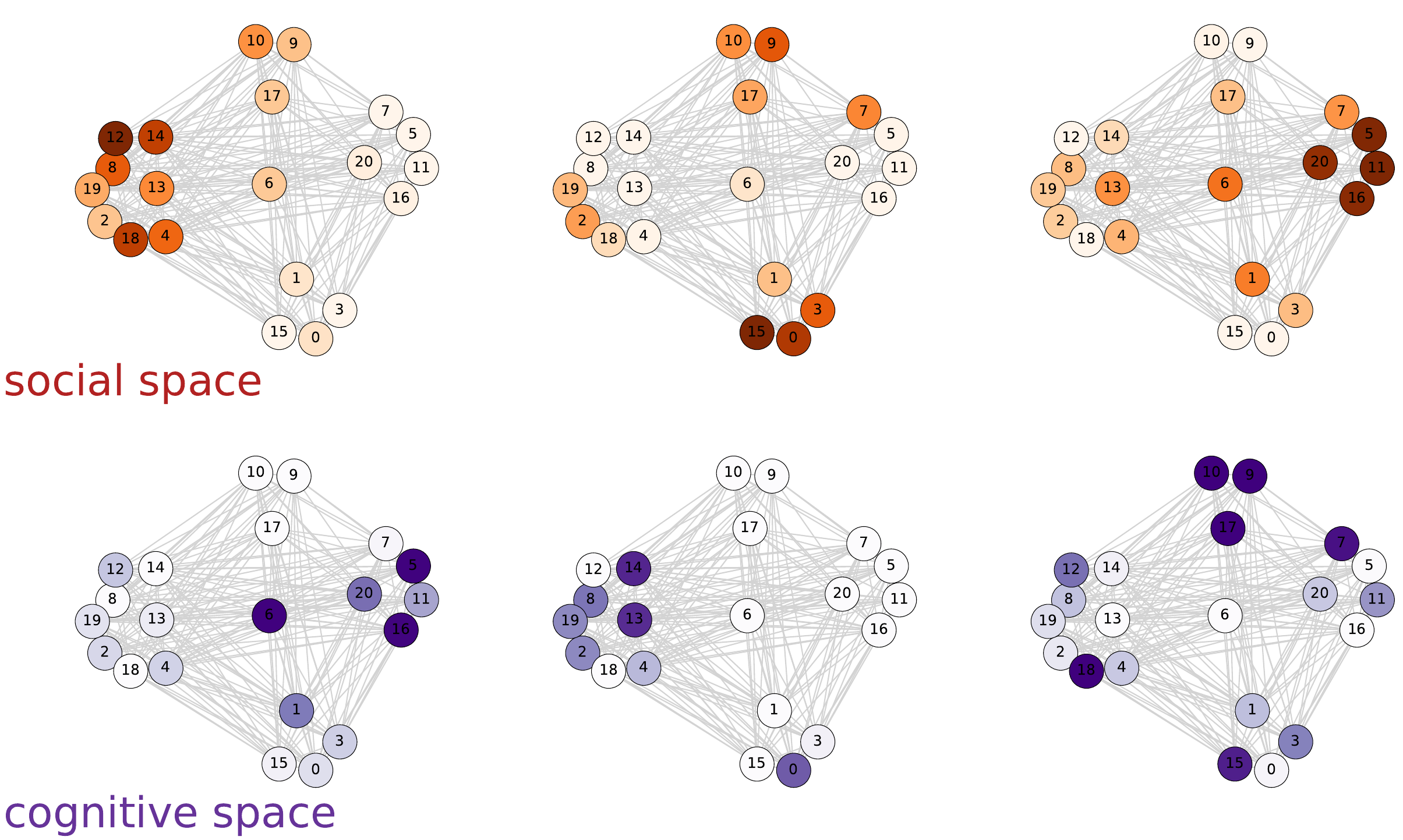}
    \caption{The latent social and cognitive spaces in the high tech firm from \protect\citet{krackhardt1987}, identified by estimating a cognitively dependent \acs{ntd} of the advice \acs{css} with $K=C=3$. The plotted network is of the network's \textit{consensus structure}, with an edge shown if at least 50\% of the network perceived its existence. Each node's position is determined by the departmental affiliation and hierarchy structure of the firm, where the person in the middle is the president, persons 1, 3, 17, and 20 are vice presidents, and the rest are supervisors. Each node is colored according to its proportional membership to each group, where a darker color denotes more proportional membership. We see that persons 5 and 16 belong mostly to the same cognitive space as the president, persons 0 and 13 belong mostly to the same cognitive space as person 14, and everyone else belongs to the third cognitive space.}
    \label{fig:cogsoc} 
\end{figure}

We inspect the latent spaces identified by this \acs{ntd} in \Cref{fig:cogsoc}. We see that the groupings of the social space of the advice network can be mostly attributed according to departmental affiliation within the firm: the three social groups mostly correspond to the left, bottom, and right, departmental groupings (visualized here in separate clusters), whereas the president (person 6) doesn't strongly belong to any of the three groups. The exception to this is the departmental grouping that we view in the upper center of the network  visualization, containing nodes 9, 10, and 17, which doesn't clearly belong to any of the three social groups. The cognitive space, however, seems to group employees according to different attributes.

\begin{figure}
    \centering
    \includegraphics[width = .85\textwidth]{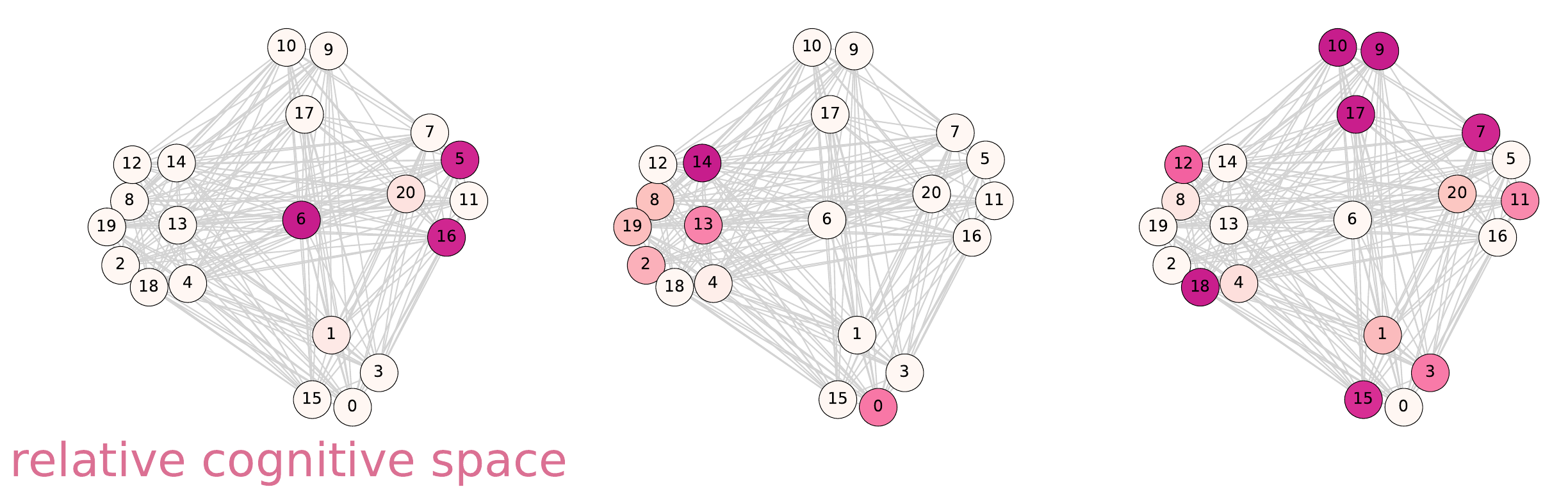}
    \caption{The latent cognitive space of the \protect\citet{krackhardt1987} advice \acs{css}, rewritten relative to the relational schema of the president of the company, the supervisor we refer to as person 14, and person 10. Each node is colored according to its proportional membership to each cognitive group, where dark pink denotes more membership. Note that, because this plot shows the cognitive membership of each node \textit{relative} to persons 6, 14, and 10, person 6 (the president) has his entire membership in the first cognitive group, and persons 14 and 10 have their entire membership in the second and third cognitive groups, respectively.}
    \label{fig:relcog}
\end{figure}

As is, these cognitive groupings aren't entirely interpretable. To further interpret the identified cognitive space, we can rewrite the $\Y$ factor matrix and the core tensor in the basis of $C=3$ individuals in the network. We choose the three individuals according to the heuristic proposed in \citet{aguiar2022factor}, where the aim is to choose $C$ layers such that the corresponding rows of $\Y$ are (nearly) linearly independent. Doing so, we identify person 6, person 14, and person 10 as the reference perspectives to consider. We transform $\Y$ such that the $C=3$ frontal slices of $\G$ correspond exactly to the affinity matrix from which persons 6, 14, and 10 generate their perceptions, respectively. Thus, the $6th$, $14th$, and $10th$ rows of the transformed $\Y^*$ matrix will be $[1,0,0]$, $[0,1, 0]$, and $[0,0,1]$, respectively, and all other rows will represent each individual's relational schema \textit{relative} to these three people. We inspect this transformed $\Y^*$ matrix, which identifies the \textit{relative cognitive space}, by plotting each individual's proportional cognitive membership in \Cref{fig:relcog}.

\include{split_LRT_table}
Whereas we know that person 6 is the president of the firm, all we nominally know about person 14 is that he is a supervisor of a department, and that person 10 is a long-standing employee. Digging further into the original analysis of this dataset, however, we find that person 14 is identified as having a particularly interesting perception of his advice network (the friendship network is not discussed). Quoting from \citet{krackhardt1987} (emphasis our own),

\begin{quote}
    \textit{``\ldots \textbf{his own perception is that he is very active in the network}: advice is sought from him by 12 people; he actively seeks advice from 20 others; and he is on the crossroads of this network as evidence by his 81.15 betweenness score. \textbf{This self-evaluation is not shared by his coworkers} \ldots \textbf{Only three of the 12 indegrees he claims to have are confirmed} \ldots Also, only nine of his 20 outdegree nominations are confirmed \ldots And finally, his betweenness in the LAS just about disappears (betweenness = $0.70$). The Consensus Structure reveals that \textbf{people generally think that no one approaches him for advice, that he goes to only five people for advice, and that he is not in between any other pair of people.}''}
\end{quote}

Thus, the relational schema that the 21 different individuals in this firm rely upon when recalling the advice network can be concisely described by the relational schema of just three distinct people: the president of the firm, this supervisor with an overly optimistic view of his surrounding advice network, and someone who seems to represent the remaining employees. We continue to analyze this network and these interpretations in the next section, as we explore and contrast the social and cognitive spaces of the friendship network. 

\subsection{Krackhardt Friendship CSS}\label{subsec:krackf}
We turn now to analyze the same firm as above, with the \acs{css} now representing perceptions of the friendship relationships. In modeling this \acs{css}, the test-AUC for different model specifications suggests using $K=3$, $C=5$ as preferable to $K=3$, $C=3$, what we used to analyse the advice network. The former has a much lower test-AUC (see \Cref{fig:xval}, right). We see that for these values of $K=3$ and $C=5$, the standard \acs{lrt} fails to reject the null hypothesis that the \acs{css} is explained by the simpler cognitively dependent \acs{ntd}. 

\begin{figure}
    \centering
    \includegraphics[width = .95\textwidth]{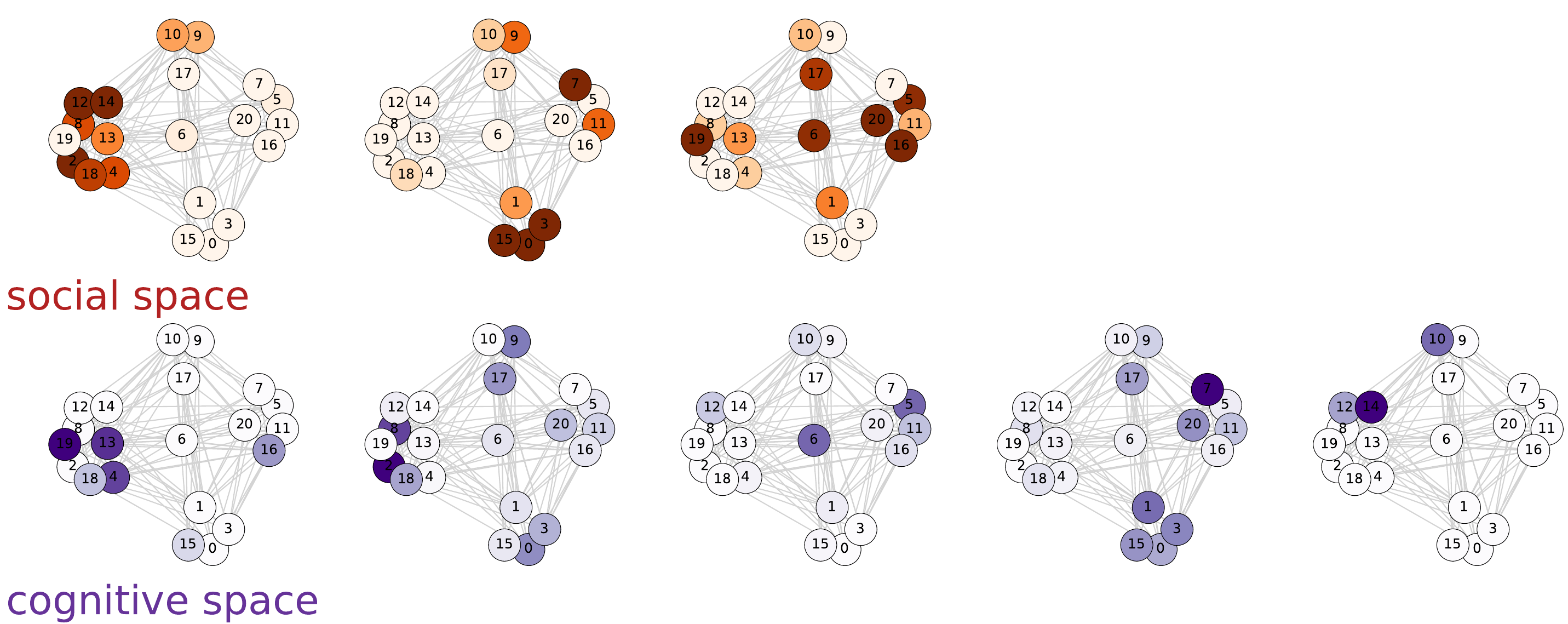}
    \caption{The latent social and cognitive spaces in the friendship \acs{css} of the high tech firm from \protect\citet{krackhardt1987}, identified by estimating a cognitively dependent \acs{ntd}.}
    \label{fig:krackf_cogsoc}
\end{figure}

We inspect the social and cognitive spaces with this model specification in \Cref{fig:krackf_cogsoc}. We observe that this decomposition identifies a similar three-dimensional social space to that identified in the advice \acs{css} above. Again, the three social dimensions mostly identify three different department affiliations within the firm. Differently from the advice network, however, we see that the president (person 6) almost entirely belongs to the third social space, whereas in the decomposition of the advice \acs{css}, his social membership was mostly spread across all three groups. Additionally, we see that person 7's social membership does not correspond to his departmental affiliation, but to that of another department. Whereas we lack ethnographic data that might support these friendship alliances, it is curious to wonder how and why these two people belong to different social groups when considering relationships defined by friendship as opposed to advice.

In comparing the identified cognitive spaces in the advice and friendship \acs{csspl} for this firm, we see that, again, person 6 and person 5 belong to the same cognitive space. We also observe again that person 14 is identified as belonging to a different cognitive space than most of the others. Although the friendship \acs{css} is not discussed in \citet{krackhardt1987}, it is reasonable to assume that person 14 might have been just as overly optimistic in reporting his perception of the friendship network as he was in the advice network. Why is his relational schema, which sets him apart from many others in the firm, so notably different from that of his colleagues? While there are any number of reasons that could describe this (internal firm politics, his socioeconomic status, his upbringing), it's also possible that he is singled out by the \acs{ntd} because of the noise that he contributed to the \acs{css} by exaggerating connections in his perceived network. In this sense, the \acs{ntd} has the potential to identify a cleaner model of the data, if we were to remove the noisy perceptions. 

We see that using the \acs{ntd} to understand the advice and friendship \acs{csspl} gives us a much cleaner and richer framework for analyzing these social networks. The original data analysis of this firm was able to identify person 14 as someone who had a noticeably abnormal perception of his social network by comparing his perception of the advice network to an aggregation of his coworkers' perceptions. The \acs{ntd}, however, and the sociological theories that it builds upon and operationalizes, gives us the vocabulary to say that he has a different \textit{relational schema} than most of his coworkers. Likewise, for the advice network, we are able to identify two other groups with shared relational schema, as well as visualize the employees who share relational schema across the groups (for instance, person 8 partially shares the relational schema of both person 14 and person 10). Furthermore, we are able to visualize how both the social and cognitive spaces change when considering the advice network in contrast to the friendship network.

A clear limitation of this present analysis is the lack of data about the individuals in this organization: each person is a manager in a high tech firm in the pacific northwest; we don't know much about the internal dynamics of the company; each person is a male aged between 27 and 59. According to \citet{howard1994social}, ``Gender, race, and age are social systems of differentiation that are especially prone to cognitive categorization.'' In this present dataset, where each individual has the same gender and race, we are unable to confirm or refute whether the latent cognitive space identifies these relational schema. 

\begin{figure}
    \centering
    \includegraphics[width = 0.5\textwidth]{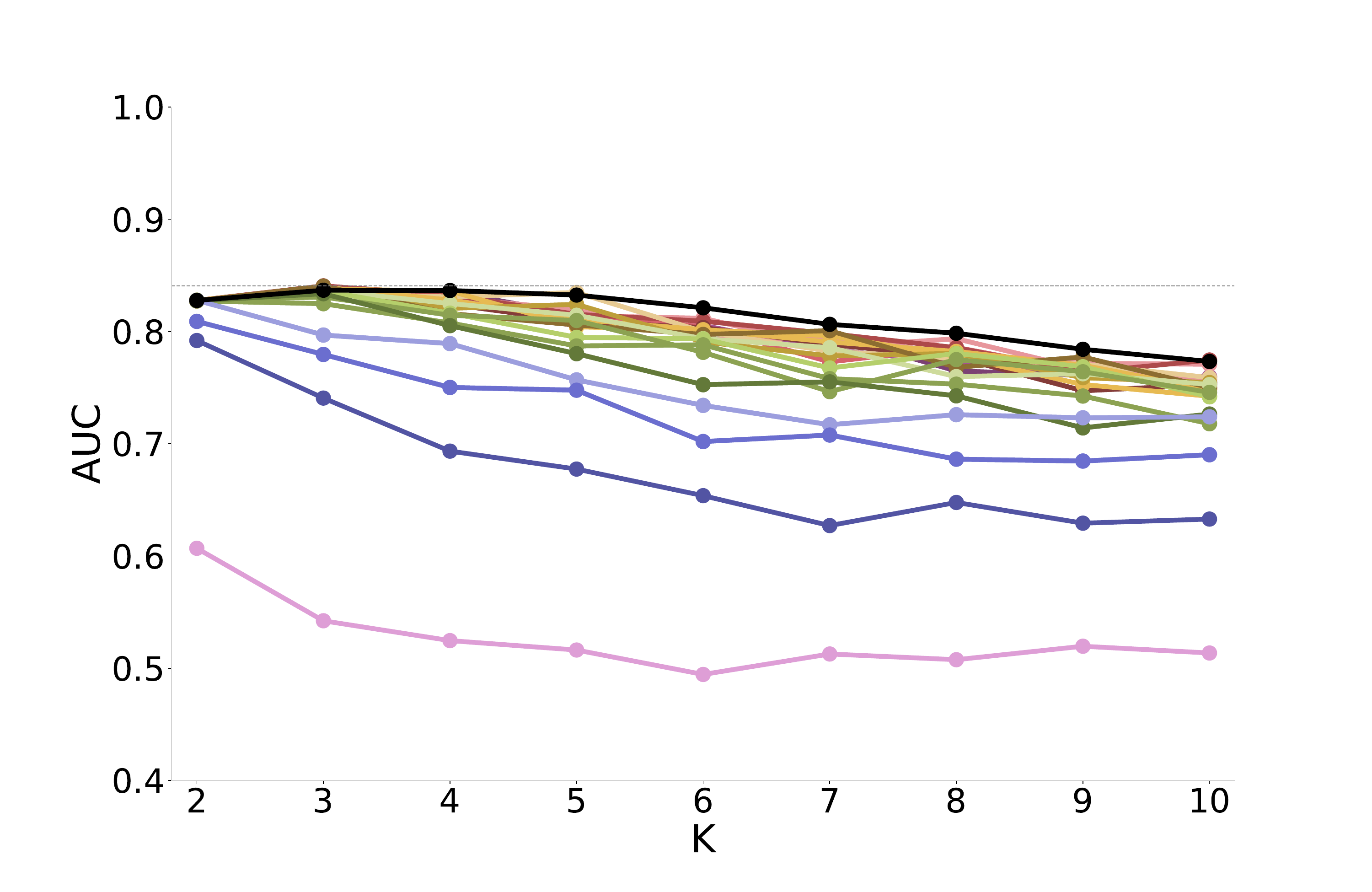}\includegraphics[width = 0.5\textwidth]{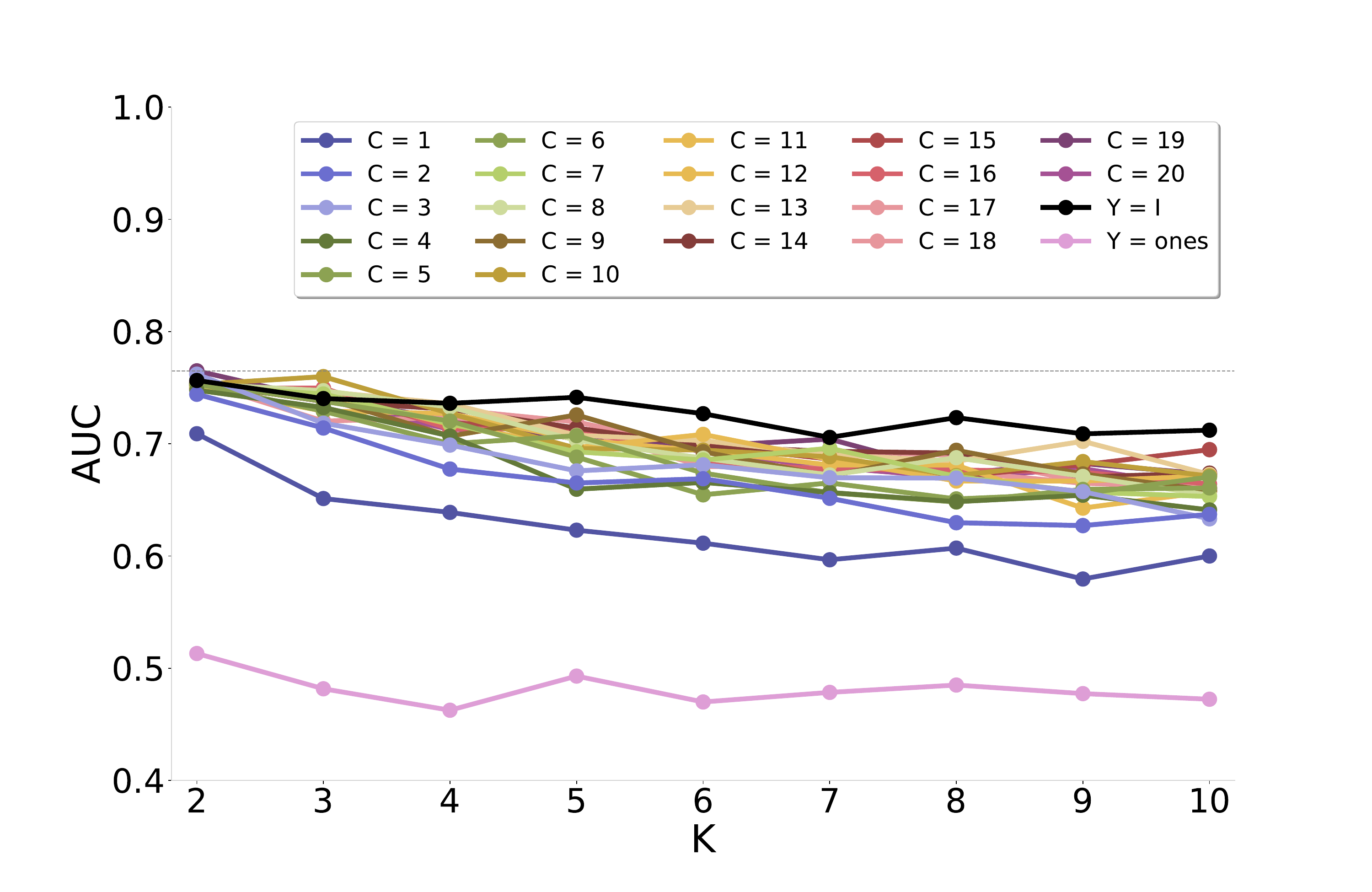}
    \caption{The test-AUC averaged across a tubular fivefold \acs{xval} task for the Hunter friendship week one (left) and week six (right) \acs{css} datasets. We choose to examine the social and cognitive factor matrices of both \acs{css} datasets corresponding to the cognitively dependent \acs{ntd} with $K = 2$ and $C=3$.}
    \label{fig:hunterxval}
\end{figure}

\subsection{Hunter Friendship CSS}\label{subsec:hunter}
We focus the analysis of this rich dataset by contrasting the identified social and cognitive spaces of week one and week six of the longitudinal study.

\paragraph{Week One.} When considering the predictive power of various \acs{ntd} models of the week one \acs{css}, we see that the cognitively dependent \acs{ntd} with $K=2$ and $C=3$ performs just as well as the cognitively independent \acs{ntd} (see \Cref{fig:hunterxval}, left). However, the \acs{lrt} comparing these two models fails to reject the smaller, cognitively dependent, model (see \Cref{table:LRT}). Considering this alongside the observation that the two models have nearly identical predictive power and similar log-likelihoods, we choose to examine the cognitively dependent model further. 

\begin{figure}
    \centering
    \includegraphics[width = .7\textwidth]{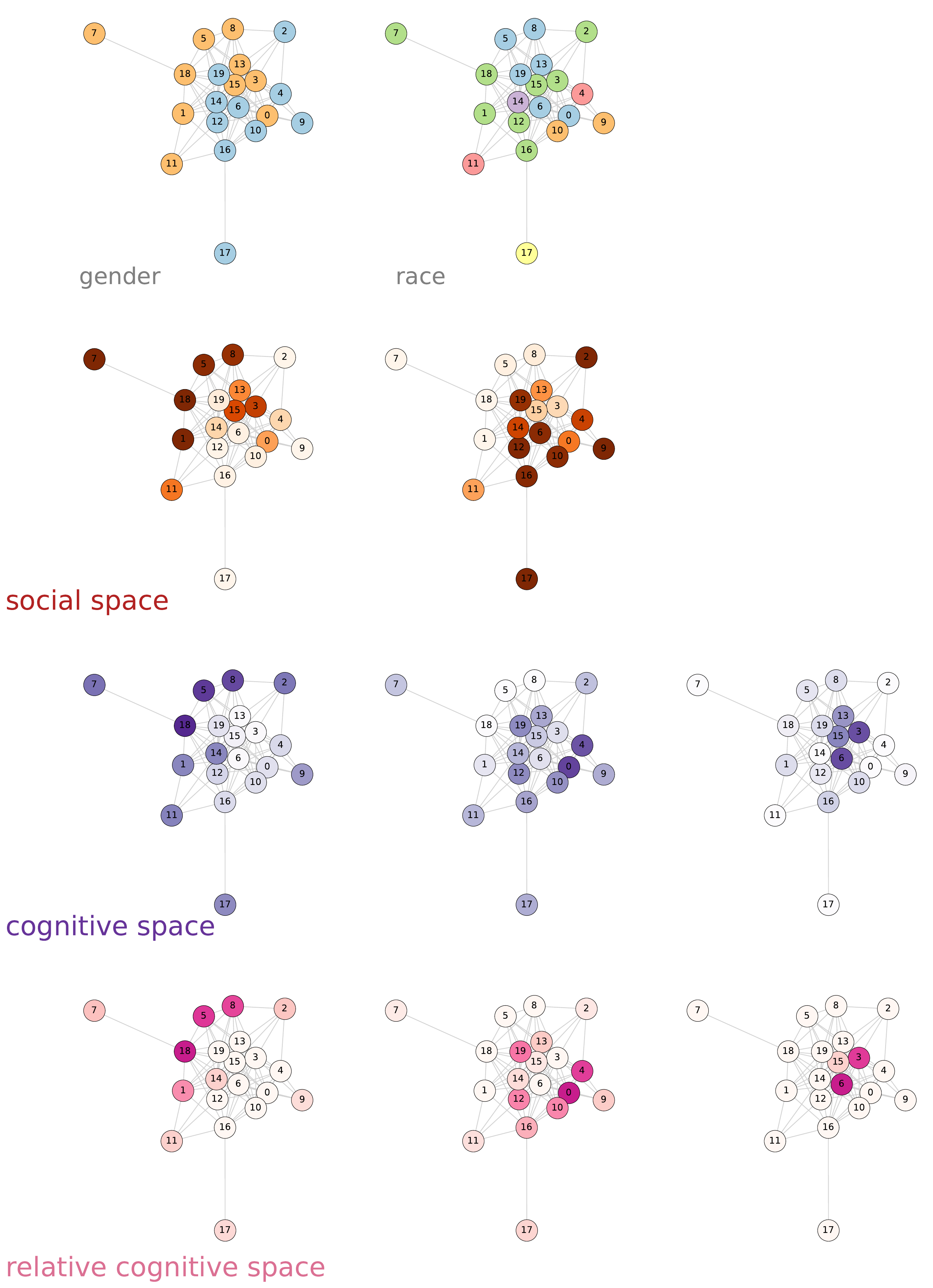}
    \caption{The latent social and cognitive spaces in week one of the Hunter Friendship \acs{css}, identified by estimating a cognitively dependent \acs{ntd} with $K=2$ and $C=3$. In the first row we plot the gender and race identifiers for each of the 20 students and in the last row we plot the cognitive space relative to the relational schema of persons 0, 6, and 18. Note that both the social and cognitive spaces in week one align well with the self-identified gender of each student. The plotted network is of the network's \textit{locally aggregated structure}, with an edge shown from node $i$ to $j$ if node $i$ perceived its existence.}
    \label{fig:hunterf0_cogsoc}
\end{figure}

We inspect the identified social and cognitive space, as well as the relative cognitive space, in \Cref{fig:hunterf0_cogsoc}. Notably, we see that during week one of the course, the students' social space can largely be described by their self-identified gender. Similarly, the identified cognitive spaces are well aligned with gender identity, where we see the first group is mostly one gender and the second group is mostly another gender. Interestingly, the third identified cognitive group, which mostly includes students 3 and 6, has students of both gender identities. While we do not have enough data at hand to understand why these students have different relational schema from the others with their same gender identity, our analysis provides an entry point for further analysis.

\begin{figure}
    \centering
    \includegraphics[width = .85\textwidth]{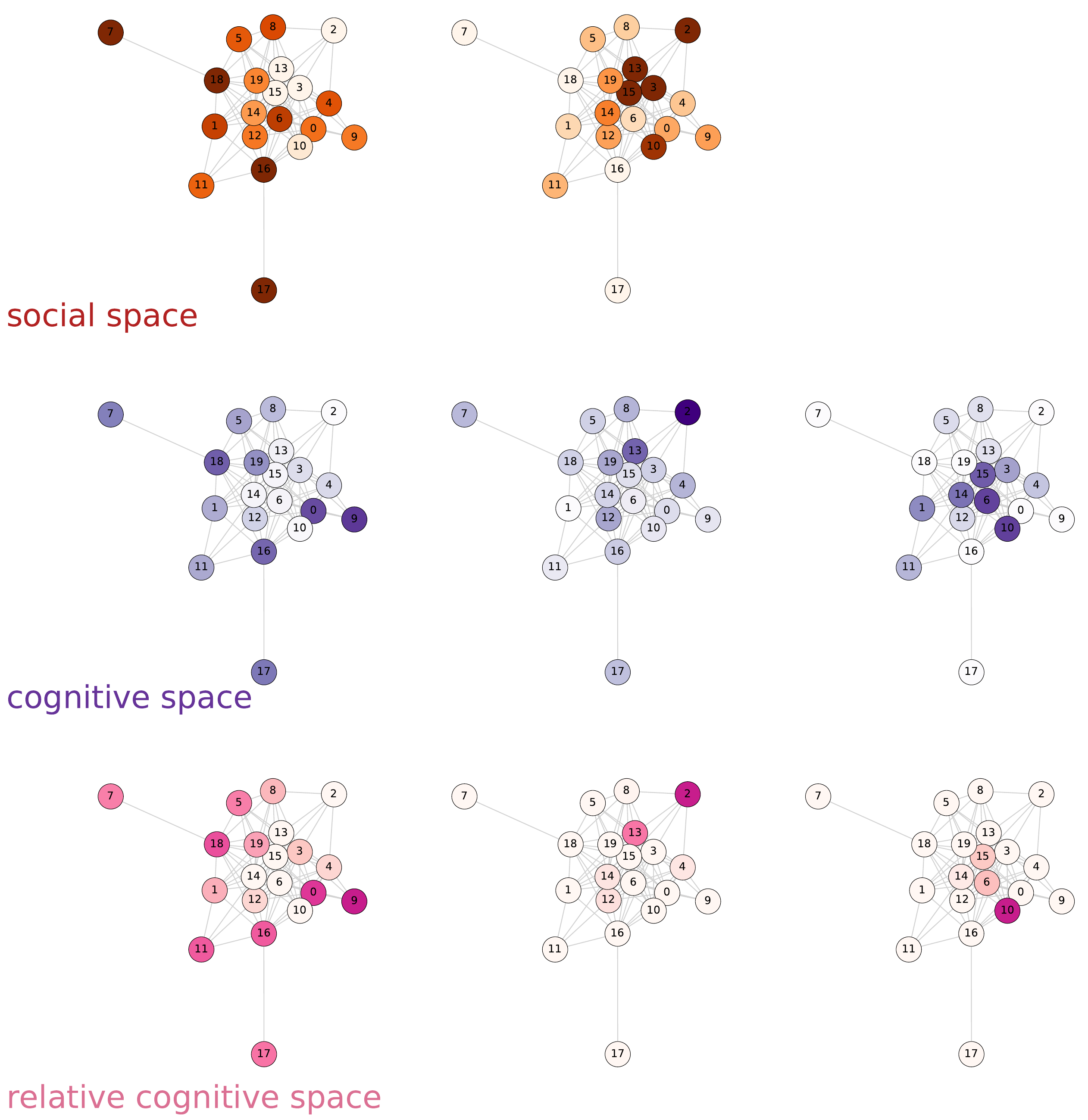}
    \caption{The latent social and cognitive spaces in the last week of the college leadership course friendship network from \protect\citet{hunter2019}, identified by estimating a cognitively dependent \acs{ntd} with $K=2$ and $C=3$. In the last row we also plot the cognitive space relative to students 9, 2, and 10. To visualize the differences in the cognitive spaces of these students, see \Cref{apx:vis_space}.}
    \label{fig:huntera_cogsoc}
\end{figure}
\paragraph{Week Six.} The test-AUC for different model specifications describing the last week of the friendship \acs{css} lead us to inspect the \acs{ntd} with $K=2$ and $C=3$ (see \Cref{fig:hunterxval}, right). With this model choice we fail to reject the null hypothesis that the advice \acs{css} was generated from a cognitively dependent \acs{ntd}, and we explore this dependent \acs{ntd} in further detail. \Cref{fig:huntera_cogsoc} gives visualizations of the identified social and cognitive spaces. 

The social spaces that are identified in the \acs{ntd} estimation are not easily explained given the metadata we have (see \Cref{fig:huntermeta} in \Cref{apdx:meta} for visualizations of available student attributes, including study groups, undergraduate institutions, and declared majors). Even without any external context about the students in each of the identified social and cognitive spaces in week six, the longitudinal nature of this dataset provides unique opportunities for other rich insights. Notably, we can learn a lot about both the students and the course by comparing these spaces to those identified in week one. We first observe that the identified social and cognitive spaces no longer correspond with the self-reported gender identity of the students. Secondly, we note that students 3 and 6, who shared relational schema in the first week, now have different relational schema, with person 3 mostly belonging to the first cognitive space and person 6 mostly belonging to the last cognitive space. While there are many possible interpretations for the social and cognitive spaces identified in the last week (see \Cref{apdx:meta} for one interpretation), the most valuable observations would come from a contextual analysis of these findings done directly by the researchers \citep[see, for example,][wherein the author compares the findings of a community detection algorithm with her own ethnographic observations of the study participants.]{carley1986approach}. To develop an understanding of the differences between the cognitive spaces identified in the last week of this \acs{css}, we consider two different visualizations in \Cref{apx:vis_space} which display different ways to understand the differences in the generative processes identified by each cognitive space.

Overall, the above observations have the potential to say a lot about this classroom setting, the instructional materials, and the students enrolled in this course. The course began with a relatively clear divide along gender lines, both in the social environment and the students' relational schema. Again, we quote \citet{howard1994social} in her theory of social cognition wherein she notes that gender is one of the ``social systems of differentiation that are especially prone to cognitive categorization.'' Furthermore, \citeauthor{howard1994social} offers additional theory about \textit{why} these initial relational schema occurred along gender lines, writing that ``systems of classification must be relatively simple in order to provide a beginning place for interaction. This interactional requirement generalizes the principle of cognitive efficiency: interaction is easier if there are fewer cognitive distinctions to consider \ldots the constant use, in interaction, of dichotomies such as gender may help to keep them simple.''

Considering the last week of the course, however, after 6 weeks of instructional material, both the social and cognitive spaces became more complicated across identifiers. Whereas the students were prone to this easy heuristic of cognitive categorization\textemdash gender\textemdash in the first week of the course, we see richer and more complicated relational schema emerge by the end of the study. Again, in \citeauthor{howard1994social}'s \citeyearpar{howard1994social} introduction of the theory of social cognition she writes, ``if schemas are to be sustained and reproduced over time \ldots they must be validated by the accumulation of resources that their enactment engenders. Schemas and resources constitute structure only when they imply and sustain each other over time.'' It could be argued that this departure from a gender-informed schema indicates the success of this particular course: that the students' initial schema was not validated and engendered.

%% file: split_LRT_table.tex
\begin{table}
\centering
\begin{tabular}{l|l|l}
\multicolumn{1}{c|}{\textbf{Dataset}}                                                  & \multicolumn{1}{c|}{\textbf{Test}}                                                  & \multicolumn{1}{c}{\textbf{split-LRT Determination}}                                                  \\ \hline
\multirow{3}{*}{Krackhardt Advice}                                                     & \begin{tabular}[c]{@{}l@{}}$H_0$: Redundant \\ $H_1$: $C = 2$\end{tabular}      & \multicolumn{1}{l}{reject $H_0$}             \\ \cline{2-3}   
& \begin{tabular}[c]{@{}l@{}}$H_0$: Dependent $K=C=3$\\ $H_1$: Independent\end{tabular} & \multicolumn{1}{l}{fail to reject $H_0$}                                                                           \\ \cline{2-3} 
 & \begin{tabular}[c]{@{}l@{}}$H_0$: SCA $K=C=3$\\ $H_1$: Dependent $K=C=3$\end{tabular} & \multicolumn{1}{l}{reject $H_0$}             \\ \hline
\multirow{3}{*}{Krackhardt Friendship}                                                 & \begin{tabular}[c]{@{}l@{}}$H_0$: Redundant \\ $H_1$: $C = 2$\end{tabular}          & \multicolumn{1}{l}{fail to reject $H_0$}             \\ \cline{2-3} 
& \begin{tabular}[c]{@{}l@{}}$H_0$: Dependent $K=3, C=5$\\ $H_1$: Independent\end{tabular}  & \multicolumn{1}{l}{fail to reject $H_0$} \\ \cline{2-3} 
 & \begin{tabular}[c]{@{}l@{}}$H_0$: SCA $K=C=3$\\ $H_1$: Dependent $K=3, C=5$\end{tabular}    & \multicolumn{1}{l}{\textbf{fail to reject $H_0$}}      \\ \hline
\multirow{3}{*}{\begin{tabular}[c]{@{}l@{}}Hunter Friendship,\\ Week One\end{tabular}} & \begin{tabular}[c]{@{}l@{}}$H_0$: Redundant \\ $H_1$: $C = 2$\end{tabular}          & \multicolumn{1}{l}{reject $H_0$}             \\ \cline{2-3} 
 & \begin{tabular}[c]{@{}l@{}}$H_0$: Dependent $K=2, C=3$\\ $H_1$: Independent\end{tabular}  & \multicolumn{1}{l}{fail to reject $H_0$}         \\ \cline{2-3} 
 & \begin{tabular}[c]{@{}l@{}}$H_0$: SCA $K=C=2$\\ $H_1$: Dependent $K=2, C=3$\end{tabular}    & \multicolumn{1}{l}{reject $H_0$}            \\ \hline
\multirow{3}{*}{\begin{tabular}[c]{@{}l@{}}Hunter Friendship\\ Week Six\end{tabular}}  & \begin{tabular}[c]{@{}l@{}}$H_0$: Redundant \\ $H_1$: Dependent $C = 2$\end{tabular}          & \multicolumn{1}{l}{reject $H_0$}             \\ \cline{2-3} 
 & \begin{tabular}[c]{@{}l@{}}$H_0$: Dependent $K=2, C=3$\\ $H_1$: Independent\end{tabular}  & \multicolumn{1}{l}{fail to reject $H_0$} \\ \cline{2-3} 
 & \begin{tabular}[c]{@{}l@{}}$H_0$: SCA $K=C=2$\\ $H_1$: Dependent $K=2, C=3$\end{tabular}    & \multicolumn{1}{l}{\textbf{fail to reject $H_0$}}            
\end{tabular}
\caption{The determinations from the \acs{slrt} for the Krackhardt and Hunter CSS datasets. See \Cref{apdx:lrts} for a discussion on the \acf{lrt}, for motivation for why we use the \acs{slrt} as opposed to the regular \acs{lrt}, and to compare these determinations to those of the standard \acf{lrt} . Of note is that the \acs{slrt} suggests that both the Krackhardt Friendship \acs{css} and the last week of the Hunter Friendship \acs{css} are well modeled with \acf{sca}. The other two \acs{css} datasets are better explained when we allow for the social space to differ from the cognitive space.}
\label{table:LRT}
\end{table}

%% file: conclusion.tex
From the analyses of the \acl{csspl} in this work, we conclude that individuals' \textit{perceptions} of their social networks are often just as, if not more, important than some identified ``true'' network. These perceptions are rich with information, and arguably more valuable when considered all together. By considering a \acs{css} data object as a tensor, and estimating a factor model of that data tensor, we are able to simultaneously consider each individual's relational schema\textemdash how they categorize and compress their perceptions of their social world\textemdash and how it relates to others'. 

This work highlights many exciting opportunities for future work. From a modeling perspective, it is interesting to connect how other more recent multilayer \acl{sbms} \citep[for example,][]{josephs2023nested} relate to the \acs{ntd}, and how their modeling interpretations in the context of \acs{css} datasets overlap or differ. Mapping the theory between seemingly disparate multilayer generative models is a fruitful direction for future work on both multilayer networks and their applications.

Above, we proposed the question of empirically studying and testing for \textit{\acl{sca}} in a network, and future work can be dedicated to further studying its sociological (or cognitive) implications. As we briefly discussed, we find evidence that two of the \acs{css} datasets we explore may be well explained by a model with \acl{sca} (see \Cref{table:LRT}). \textit{When} can we expect to see \acl{sca}? \textit{What} does it mean when \acl{sca} is or isn't observed in a group? Do only certain types of social relationships reflect an equivalence in social and cognitive space? A domain-informed perspective and study of these questions (perhaps with an accompanying \acs{css} dataset) which uses the tools proposed here (or other appropriate latent space models) to empirically study these specific questions would provide rich insight into the cognitive processes underlying social networks. Furthermore, although \acl{sca} is proposed to model and test for when the social and cognitive spaces are the same, there may be applications in which it is interesting to compare the differences between the identified social and the cognitive spaces when they're \textit{not} constrained to be equivalent. In such a setting, one could consider quantifying the difference between the spaces using a cluster comparison measure such as Variation of Information \citep{meila2003}.

At a technical level, as discussed in the main text as well as in \Cref{apdx:sca}, there are also several open questions regarding the optimization techniques we propose for estimating the \acl{sca} \acs{ntd}. As presented, it does not have monotonic convergence guarantees (and none of the multiplicative update algorithms for tensor factorization have any guarantees of global optimality). As such, we cannot confidently assume that the resulting estimated \acs{ntd} satisfies the maximum likelihood assumption necessary in both the standard \acs{lrt} (or \acs{slrt}, as discussed in \Cref{apdx:lrts}). For future work aiming to identify and/or test \acl{sca} in a network, efforts to improve the suggested algorithm and test would be very helpful contributions. 

As another direction for future work, we saw in the analysis of the Krackhardt \acs{css} that the estimated cognitive space identified persons 6, 10, and 14 as having notably different relational schema from one another. In \Cref{subsec:krackf} we discuss how person 14 may have been singled out due to his notably dense and optimistic perceptions of the network. Although we focused this work on the introduction of the \acs{ntd} as a tool for studying \acs{csspl}, interesting future work could further explore this question of identifying \textit{noise} in \acs{css} datasets, possibly comparing the cognitive space of the \acs{ntd} to the latent methods proposed by \citet{sewell2019latent} and \citet{debacco2023}.

In a separate direction, recent work on information diffusion has focused on how information and behaviors spread differently across different \textit{layers} of multilayer social networks, aiming to identify which types of relationships are most important to identify for purposes of influence maximization \citep{kempe2003}. This examination of information cascades in a multilayer perspective readily lends itself to interesting questions in the \acs{css} space, namely, \textit{how does information propagate across different network perspectives?} and \textit{does information seeded in different cognitive spaces spread differently?} It's well established that seeding information in different parts of a social space leads to differences in diffusion \citep{krackhardt1996structural, banerjee2013}. To go a step further, we contend that perhaps differences in cognitive space---when departing from social space, see our earlier discussion of social-cognitive agreement---are even more significant. When people share a piece of information, they do not have an omniscient view of some ground truth network: they act and share according to their perceptions. Indeed, to quote \citet{thomasthomas}, just as \citeauthor{krackhardt1987} did in his original work, ``if men define situations as real, they are real in their consequences.'' We believe factor-based analyses of \acl{css}, using the NNTuck, can meaningfully help us better understand how cognition impacts information propagation. 

Currently, \acs{css} data is collected by asking each person \textit{yes/no} questions about relationships in their surrounding network. Another interesting direction for future work is incorporating something akin to an \textit{unknown} response into the space of possible answers for \acs{css} surveys. Distinguishing between perceived relationships, educational guesses-at relationships, and pure speculation of relationships ought to have interesting impacts on how \acs{csspl} can be interpreted, including through properly adjusted \acs{ntd} models, showing how and when relational schema are most important. 

Finally, there is a obvious lack of \acs{css} data from networks consisting of more than 30 people. This is, in most part, due to the immense burden (both on the survey administrator and participant) of asking each person in a network of size $N$ to report on $N^2-N$ distinct relationships. Collecting the entire \acs{css}, however, may be unneccessary depending on the research objective. For example, if the end goal of collecting \acs{css} data is to identify the generative \acs{ntd} and the network's associated latent social and cognitive spaces, partial data may be more than sufficient. Extending the active matrix factorization for surveys work of \citet{zhang2020active} to tensors could potentially allow for efficient \acs{css} data collection of much larger networks.

We conclude by urging that there is much to be learned from continuing to view the \acs{css} as a tensor-object. We hope that the present work opens new research directions towards the aim of uncovering relational schema in social networks, and are eager for future work that continues in this pursuit. 

%% file: appendix.tex
\appendix
\appendixpage
\section{Hunter CSS metadata}\label{apdx:meta}
\begin{figure}
    \centering
    \includegraphics[width = 0.8\textwidth]{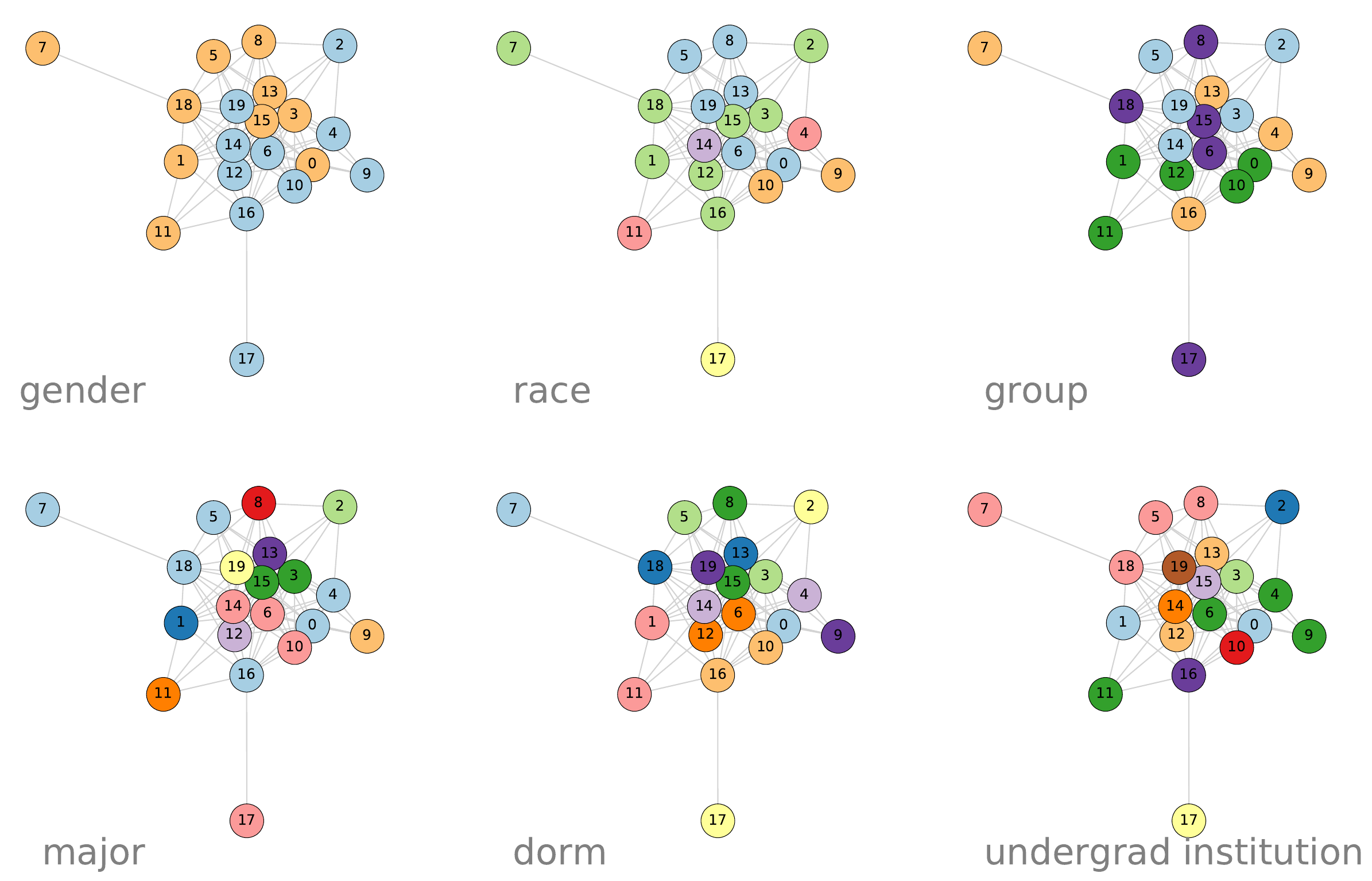}
    \caption{The metadata of the students in the leadership course studied in \protect\citet{hunter2019}. }
    \label{fig:huntermeta}
\end{figure}

We briefly examine and visualize (see \Cref{fig:huntermeta}) the available metadata of the students and classroom from the longitudinal \acs{css} study from \citet{hunter2019} which we analyze in \Cref{subsec:hunter}. The ``group'' attribute visualized in \Cref{fig:huntermeta} describes the study groups the students were assigned to at the beginning of the study. Not only were students within a group expected to do in-class exercises together, but they also participated in a Prisoner's Dilemma activity together in the middle of the course. At the end of the Prisoner's Dilemma activity, each group was asked to describe the actions of the other groups in the class. These descriptions are reported in the ``affect matrix'' in \Cref{tab:affect}. For more details, see \citet{hunter2019}. Note that Group A and Group D had, and made, very different impressions of, and on, each group. Specifically, Group A described all groups' actions with positive words like ``honest'' or ``profitable,'' whereas group D described all groups' actions with negative words like ``conflicted'' and ``sell out.'' Similarly, all other groups described Group A's actions with negative words like ``filth'' and ``self-interested,'' whereas all other groups described Group D's actions with positive words like ``loyal'' and ``altruistic.'' 

Although more context is needed to make any definitive conclusions, these differences lend more insight into the observations made in \Cref{subsec:hunter}. Specifically, in the first week, students 3 and 6 had shared schema that distinctly differed from the other students', whereas in the sixth week their schema differed (they also belong to separate social groups in both the first and sixth week). Interestingly, student 3 was in Group A and student 6 was in Group D. Again, interesting future work could examine the entire longitudinal \acs{css} dataset as a fourth-order tensor to consider \textit{when} these two students stopped sharing a relational schema, using the context and timing of the Prisoner's Dilemma activity as additional information.

\input{affect_table}

\section{Algorithmic implementation details}\label{apdx:sca}
Here, we provide more details of the algorithm used to estimate the \acs{ntd}, as discussed in \Cref{subsec:opt_det}. We also explore the choice of the optimization algorithm discussed in \Cref{subsec:SCA} to maximize the log-likelihood of the \acf{sca}~\acs{ntd}. 

\subsection{Kim and Choi's (2007) Multiplicative Updates}
For completeness, we first include the details of the algorithm we use to estimate the \acs{ntd}, as given in \citet{kimChoi}, which we refer to throughout the text as \Cref{nntuck:alg}. Note that in practice we choose the \acs{ntd} corresponding the highest log-likelihood (equivalent to the lowest \acs{kld}) over 20 random initializations using \Cref{nntuck:alg}. See the red lines in \Cref{fig:sca_alg_compare} to see the variation of \acs{kld} of this algorithm over 20 random initializations. For further discussion of this algorithm, see \citet{kimChoi} or \citet{aguiar2022factor}.
\begin{algorithm}
\caption{Multiplicative Updates for minimizing KL-Divergence in the \acs{ntd} \citep{kimChoi}}\label{MU-NTD} 
\begin{flushleft}
\textbf{Input:} $\A, K, C,$ \verb|Symmetric|, \verb|Masked|, $\ten{M}$, \verb|Independent|, \verb|Redundant|\\
\textbf{Initialize} $\U, \V \in \Gspace^{N \times K}, \Y \in \Gspace^{L \times C},$ and $\G \in \Gspace^{\Gdims}$ to have random, nonnegative entries.\\
if \verb|Symmetric|: $\V \leftarrow \U$, $\mat{G}_\ell \leftarrow \mat{G}_\ell^\top \mat{G}_\ell$ for $\ell = 1,\dots, C$, and skip each $\V$ update step below.\\
if \verb|Independent|: $\Y \leftarrow \mat{I}$ and skip each $\Y$ update step below.\\
if \verb|Redundant|: $\Y \leftarrow$ \verb|ones(C)| and skip each $\Y$ update step below.\\
 Initialize $\Ahat = \G \times_1 \U \times_2 \V \times_3 \Y.$\\
if not \verb|Masked|: $\ten{M} = 1^{\Adims}$ \vspace{.5cm}\\ 
while $\frac{KL(\A || \Ahat_{t})- KL(\A || \Ahat_{t-1})}{KL(\A || \Ahat_{t})}< $ \verb|rel_tol|:
\begin{align*}
    \U &\leftarrow \U \circ \frac{[\mat{M}_{(1)} \circ \mat{A}\uf{1}/ \mat{\hat{A}}\uf{1}] [\G\times_2 \V\times_3\Y]\uf{1}^\top}{\mat{M}\uf{1}[\G\times_2 \V\times_3 \Y]\uf{1}^\top} \\
    \V &\leftarrow \V \circ \frac{[\mat{M}\uf{2} \circ \mat{A}\uf{2}/ \mat{\hat{A}}\uf{2}] [\G\times_1 \U\times_3\Y]\uf{2}^\top}{\mat{M}\uf{2}[\G\times_1 \U\times_3 \Y]\uf{2}^\top} \\
    \Y &\leftarrow \Y \circ \frac{[\mat{M}\uf{3} \circ \mat{A}\uf{3}/ \mat{\hat{A}}\uf{3}] [\G\times_1 \U\times_2\V]\uf{3}^\top}{\mat{M}\uf{3}[\G\times_1 \U\times_2 \V]\uf{3}^\top} \\
    \G &\leftarrow \G \circ \frac{[\ten{M} \circ \A/ \Ahat] \times_1  \U^\top\times_2\V^\top \times_3 \Y^\top}{\ten{M}\times_1 \U^\top \times_2 \V^\top \times_3 \Y^\top} \\
    \Ahat &\leftarrow \G \times_1 \U \by_2 \V \by_3 \Y
\end{align*}
\textbf{Return} $\U, \V, \Y, \G$.
\end{flushleft}
\label{nntuck:alg}
\end{algorithm}

\subsection{Comparison of SCA algorithms}
\Cref{nntuck:SCAalg} is only one of many possibilities for how one could alter the algorithm from \citet{kimChoi} to accommodate the structure of \acl{sca}. Here we compare our choice with two other options. For each description, we assume that $\Y$ and $\U$ are equivalent upon initialization. 
\begin{algorithm}
\caption{Multiplicative Updates for minimizing KL-Divergence in the \acl{sca} \acs{ntd} }\label{SCAMUNTD} 
\begin{flushleft}
\textbf{Input:} $\A, K, C,$ \verb|Symmetric|, \verb|Masked|, $\ten{M}$, \verb|Independent|, \verb|Redundant|\\
\textbf{Initialize} $\Y, \V \in \Gspace^{N \times K}$ and $\G \in \Gspace^{\Gdims}$ to have random, nonnegative entries.\\ 
Initialize $\U = \Y$.\\
Initialize $\Ahat = \G \times_1 \U \times_2 \V \times_3 \Y.$\\
Initialize \verb|min_nntuck| $= (\U, \V, \Y, \G)$ \\
Initialize \verb|min_KL| $= KL(\A || \Ahat)$\\
if not \verb|Masked|: $\ten{M} = 1^{\Adims}$ \vspace{.5cm} \\
while $\frac{KL(\A || \Ahat_{t})- KL(\A || \Ahat_{t-1})}{KL(\A || \Ahat_{t})}< $ \verb|rel_tol|:
\begin{align*}
    &\V \leftarrow \V \circ \frac{[\mat{M}\uf{2} \circ \mat{A}\uf{2}/ \mat{\hat{A}}\uf{2}] [\G\times_1 \U\times_3\Y]\uf{2}^\top}{\mat{M}\uf{2}[\G\times_1 \U\times_3 \Y]\uf{2}^\top} \\
    &\U \leftarrow \U \circ \frac{[\mat{M}_{(1)} \circ \mat{A}\uf{1}/ \mat{\hat{A}}\uf{1}] [\G\times_2 \V\times_3\Y]\uf{1}^\top}{\mat{M}\uf{1}[\G\times_2 \V\times_3 \Y]\uf{1}^\top} \\
    &\Y \leftarrow \Y \circ \frac{[\mat{M}\uf{3} \circ \mat{A}\uf{3}/ \mat{\hat{A}}\uf{3}] [\G\times_1 \U\times_2\V]\uf{3}^\top}{\mat{M}\uf{3}[\G\times_1 \U\times_2 \V]\uf{3}^\top} \\
    &\U , \Y \leftarrow (\U + \Y)/2 \\
    &\G \leftarrow \G \circ \frac{[\ten{M} \circ \A/ \Ahat] \times_1  \U^\top\times_2\V^\top \times_3 \Y^\top}{\ten{M}\times_1 \U^\top \times_2 \V^\top \times_3 \Y^\top} \\
    &\Ahat_t \leftarrow \G \times_1 \U \by_2 \V \by_3 \Y\\
    &\text{\texttt{if }} KL(\A || \Ahat_t) < \text{\texttt{min\_KL:}}\\
    &\,\,\,\,\, \,\,\,\,\,\,\text{\texttt{min\_nntuck}} = (\U, \V, \Y, \G) \\
    &\,\,\,\,\, \,\,\,\,\,\,\text{\texttt{min\_KL}} = KL(\A || \Ahat)
\end{align*}
\textbf{Return} \verb|min_nntuck|.
\end{flushleft}
\label{nntuck:SCAalg}
\end{algorithm}

\paragraph{Alternative 1: Update $\Y$ and $\U$ and then set both to their average (``averaged'').} In this approach, for each multiplicative iteration, we proceed by first updating $\V$, then updating $\U$, and then updating $\Y$, all according to the same multiplicative updates as in \Cref{nntuck:alg}. After updating $\Y$ and before updating $\G$, we set both $\U$ and $\Y$ to be equal to their average. Note that this is the algorithm we discuss in \Cref{subsec:SCA} and which is detailed in \Cref{nntuck:SCAalg}.

\paragraph{Alternative 2: Enforce $\Y=\U$ after each update (``naive method'').} In this approach, we update each factor, beginning with $\V$, according to the update rules in \Cref{nntuck:alg}. After updating $\U$, we set $\Y =\U$, and after updating $\Y$, we set $\U = \Y$. We continue to update $\G$ as usual. As in \Cref{nntuck:SCAalg}, we observe empirically that the updates are frequently nonmonotonic, more often so than our chosen algorithm. As such, we keep track of and return the \acs{ntd} corresponding to that with the minimal \acs{kld} across iterations. This method has the downside that at each step, only either $\U$ or $\Y$ is being updated according to the correct gradient descent step, whereas the other factor is constrained to follow the gradient descent step of the other.

\paragraph{Alternative 3: Update $\Y$ and $\U$ using the average multiplicative update (``averaged updates'').} Here, we again begin each iteration with updating $\V$ according to the update rules in \Cref{nntuck:alg}. We find the multiplicative update for both $\U$ and $\Y$, which we call $\U_{update}$ and $\Y_{update}$, respectively, but do not yet update $\Y$ or $\U$. We then update both $\Y$ and $\U$ by multiplying each by the average of these updates, $\frac{1}{2}(\U_{update}+\Y_{update})$. $\G$ is updated as usual. Again, we observe that these updates are frequently nonmonotonic in \acs{kld}, and we keep track of and return the \acs{ntd} corresponding to that with the minimal \acs{kld} across iterations.

\begin{figure}
    \centering
    \includegraphics[width = 0.8\textwidth]{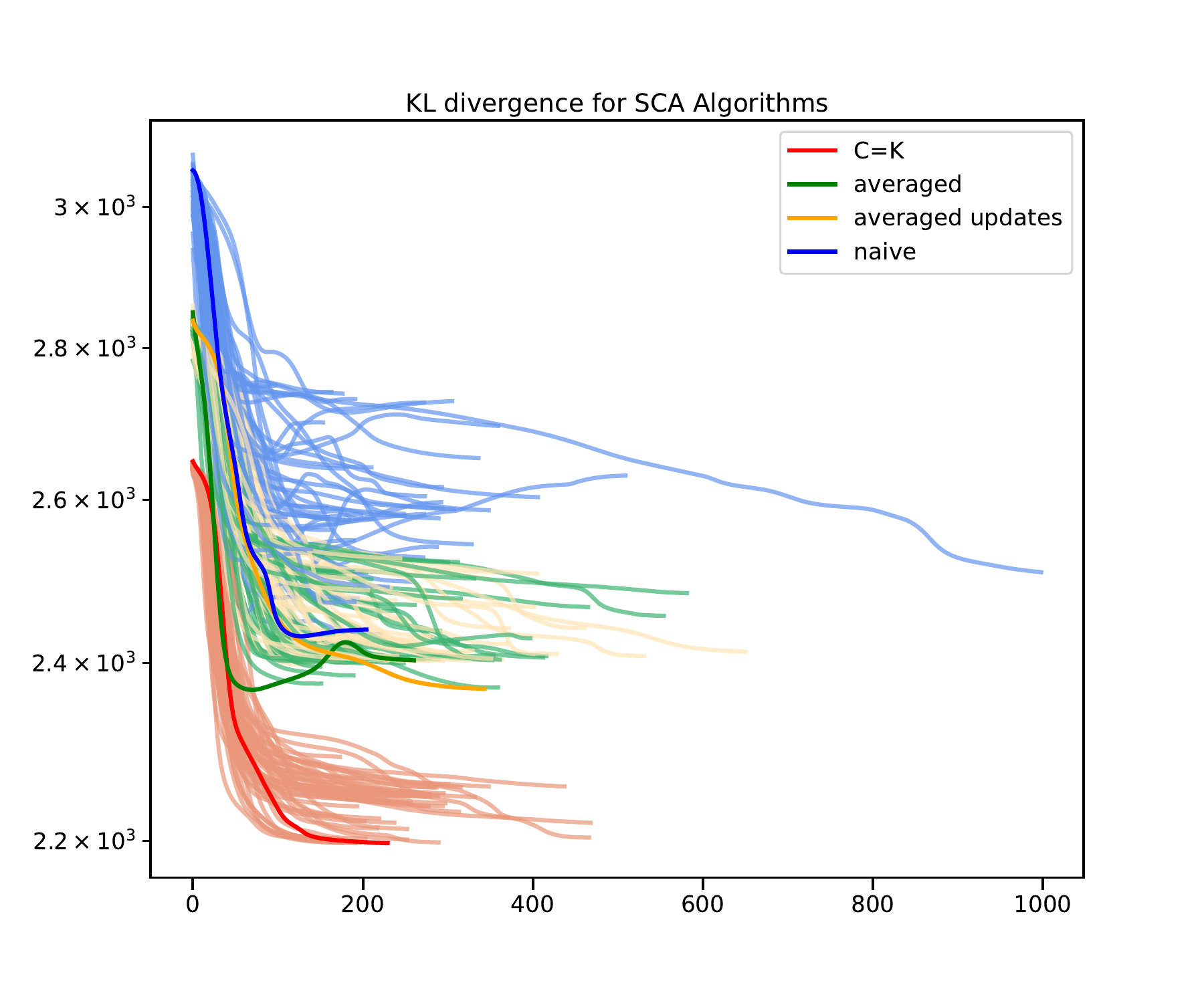}
    \caption{Comparing the \acs{kld} across iterations for 50 random initializations in three different \acs{sca} algorithms (in blue, yellow, and green), to the convergence of a dependent \acs{ntd} with $K=C$ (red) for the Krackhardt Advice \acs{css} with $K=C=3$. The comparison of these \acs{sca} algorithms to the dependent \acs{ntd} is not to compare overall difference in \acs{kld}, but rather to show how the \acs{sca} algorithms are generally nonmonotonic.}
    \label{fig:sca_alg_compare}
\end{figure}

\paragraph{Comparisons.}
 We now estimate the \acs{sca}~\acs{ntd} of the Krackhardt Advice \acs{css} with $K=C=3$ using the proposed algorithms above. In \Cref{fig:sca_alg_compare} we compare the above ``naive'' and ``averaged updates'' algorithms (in blue and yellow, respectively) with \Cref{nntuck:SCAalg} (in green). Each line shows the \acs{kld} across iterations in each algorithm for 50 random initializations. The \acs{ntd} corresponding to the highest log-likelihood (lowest \acs{kld}) across the different initializations is highlighted in bold. We also include the \acs{kld} of the unconstrained \acs{ntd} optimized using \Cref{nntuck:alg} (in red), which exhibits monotonic convergence to a local optima, whereas the other three algorithms do not (necessarily) have monotonic convergence. The performance of \Cref{nntuck:alg} can be viewed as a lower bound of sorts (but for the fact that it also doesn't have any global optimality gauarntee), since the unconstrained model is strictly more general than the constrained \acs{sca} model that the three constrained algorithms are trying to estimate.

We see that \Cref{nntuck:SCAalg} (green) generally results in a \acs{sca} \acs{ntd} with lower \acs{kld} than the other two explored algorithms, both in terms of the best-of-50 estimate and the general ensemble of estimates under different initializations.

\clearpage 

\section{Supplemental discussion of statistical tests}\label{apdx:lrts}
We now expand on the discussion of assumptions underlying statistical tests initiated in \Cref{sec:tests}. As further background, the \acf{lrt} is a statistical test of two hypotheses about the underlying generative model of a given dataset. Broadly, the \acs{lrt} compares the goodness-of-fit in terms of the maximum log-likelihood of a \textit{nested model} to that of a \textit{full model}, where the nested model is a specific instance of the full model. The \acs{lrt} compares the maximum log-likelihood of each model while considering the difference in the number of parameters between each model. The null hypothesis, then, is that both the nested and full model fit the data equally well, and thus the nested model (with less parameters) should be used. The alternate hypothesis is that the full model fits the data significantly better than the nested model, and thus the full model should be used. The standard \acs{lrt} depends on Wilks' Theorem \citep{wilks1938}, which broadly says that under the null hypothesis, as the sample size approaches infinity, the difference between the maximum likelihoods of each model will be $\chi^2$ distributed with degrees of freedom given by the difference in the number of parameters.
\include{LRT_table}
\begin{figure}
    \centering
    \includegraphics[width = 0.8\textwidth]{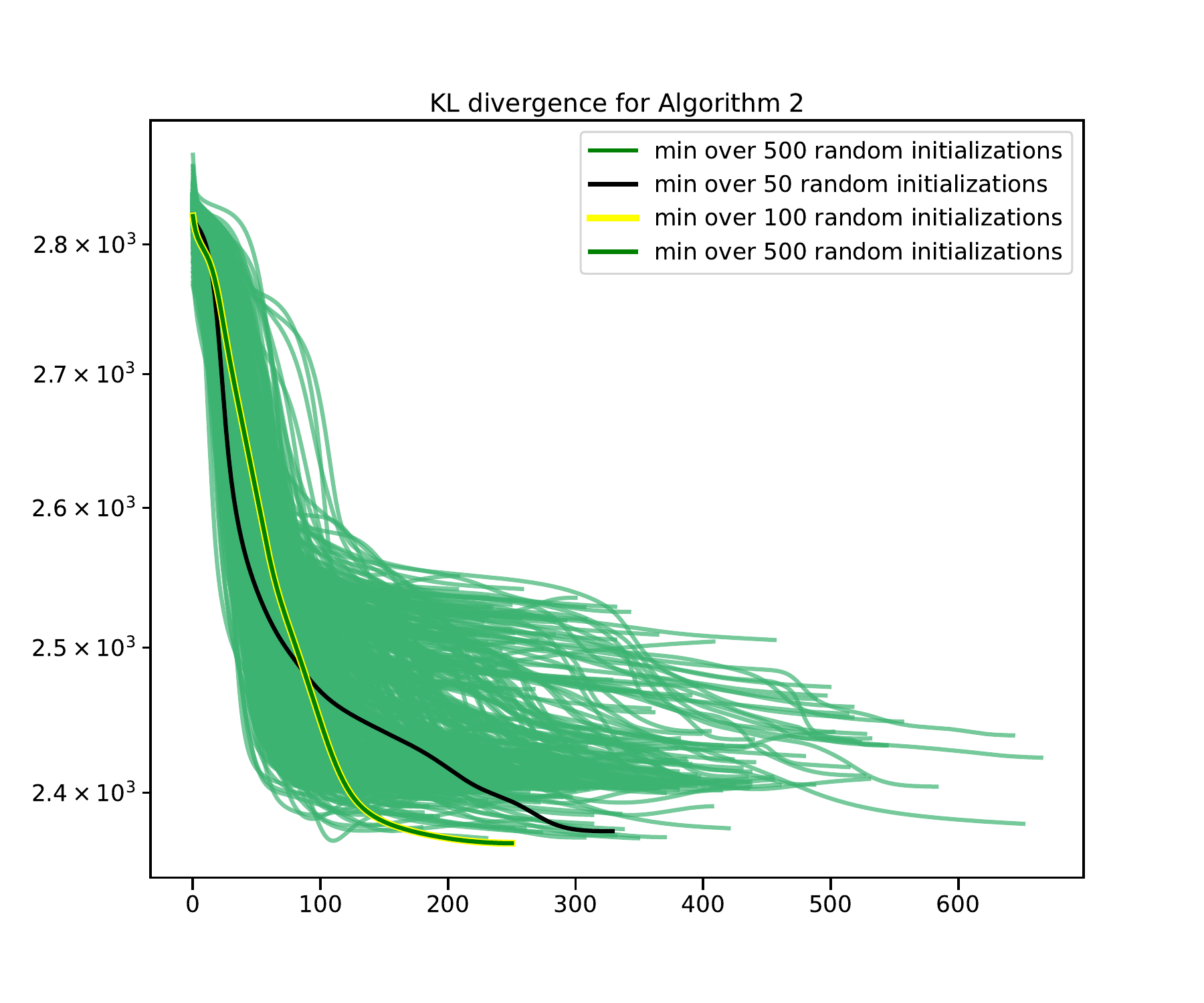}
    \caption{Comparing the \acs{kld} of \Cref{nntuck:SCAalg} across iterations for 500 random initializations. We see that the \acs{kld} of the \acs{ntd} with the minimal \acs{kld} over 50 random initializations is not much lower than the \acs{kld} of the \acs{ntd} with the minimal \acs{kld} over 500 random initializtions (which is the same as that over the first 100 random initializations). Thus, we estimate the \acs{sca}~\acs{ntd} using 50 random initializations of \Cref{nntuck:SCAalg}.}
    \label{fig:sca_numit}
\end{figure}
As introduced in the main text, the setting in which we want to statistically compare different \acs{ntd} models is not one which satisfies the necessary assumptions or regularity conditions in order for Wilks' Theorem to hold: the \acl{ntd} is neither identifiable \citep{Kolda2009, chen2020note}, nor can we be certain that our estimation of the \acs{ntd} has reached a global maximum of the log-likelihood. Thus, we instead utilize the \acs{slrt} from \citet{wasserman2020}, which does not require identifiability of the model. 

 The second shortcoming, of reaching a global maximum log-likelihood, remains troublesome for the cognitively redundant, independent, and dependent \acs{lrts} (see \Crefrange{def:lrt-ind}{def:lrt-red}), as well as the \acs{sca}~\acs{lrt} (\Cref{def:lrt-sca}). Specifically, even the \acs{slrt} requires that the model of the null hypothesis must correspond to the maximum log-likelihood. When estimating a redundant, dependent, or independent \acs{ntd}, \Cref{nntuck:alg} has guaranteed monotonic convergence to a \textit{local} minima of the \acs{kld} (equivalent to a local maxima of the log-likelihood). We attempt to overcome this local-only guarantee by choosing the \acs{ntd} with the maximal log-likelihood over many random initializations. See the red lines in \Cref{fig:sca_alg_compare} for a comparison of how the minimal \acs{kld} varies over different random initializations.

However, when estimating a \acs{sca}~\acs{ntd} we use \Cref{nntuck:SCAalg}, which is not even monotonic in \acs{kld} nor does it come with any convergence guarantees, even to a local minima. Our approach is, as discussed in the main text, to keep track of the \acs{sca}~\acs{ntd} with the highest log-likelihood over many random initializations. See \Cref{fig:sca_numit} for a comparison of how the minimal \acs{kld} varies over different random initializations when using \Cref{nntuck:SCAalg}.

Although we hope that the heuristic of utilizing multiple random initializations is able to address the problem of only having local minima convergence guarantees (or no convergence guarantees at all, in the case of using \Cref{nntuck:SCAalg} for the \acs{sca}~\acs{ntd}), we are unable to provide any assurances or certificates. As such, we urge practitioners to approach the determinations of the \acs{sca}~\acs{lrts} (both standard and split) with caution.

\section{Visualizing differences in the cognitive spaces}\label{apx:vis_space}
\begin{figure}
    \centering
    \includegraphics[width = .85\textwidth]{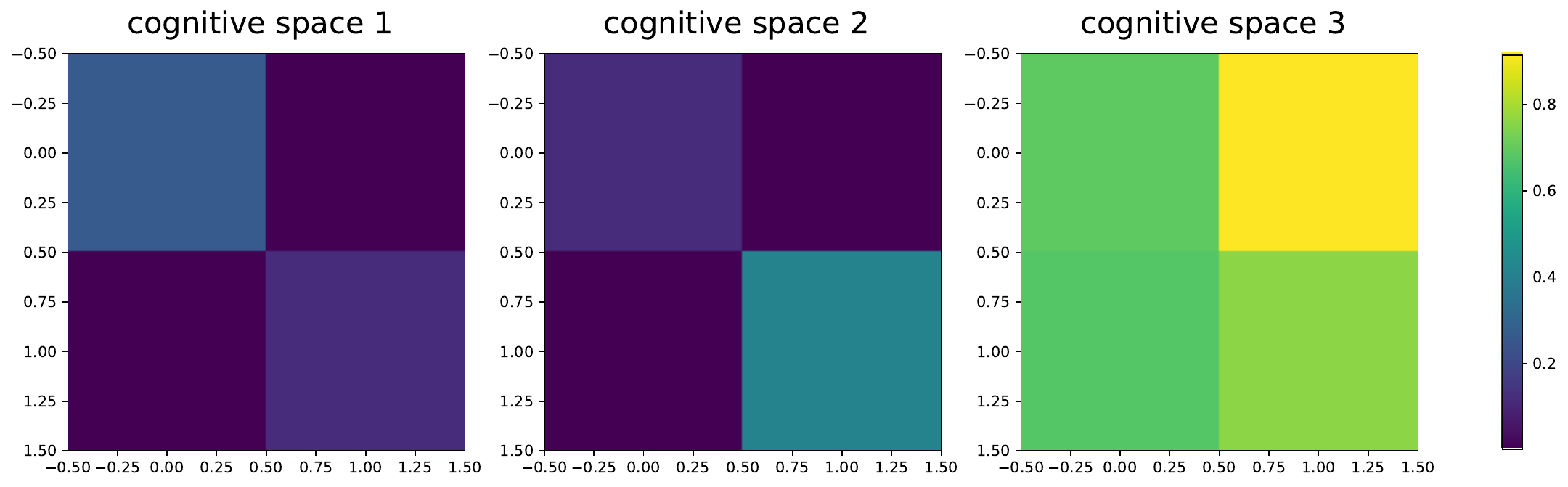}
    \includegraphics[width = .85\textwidth]{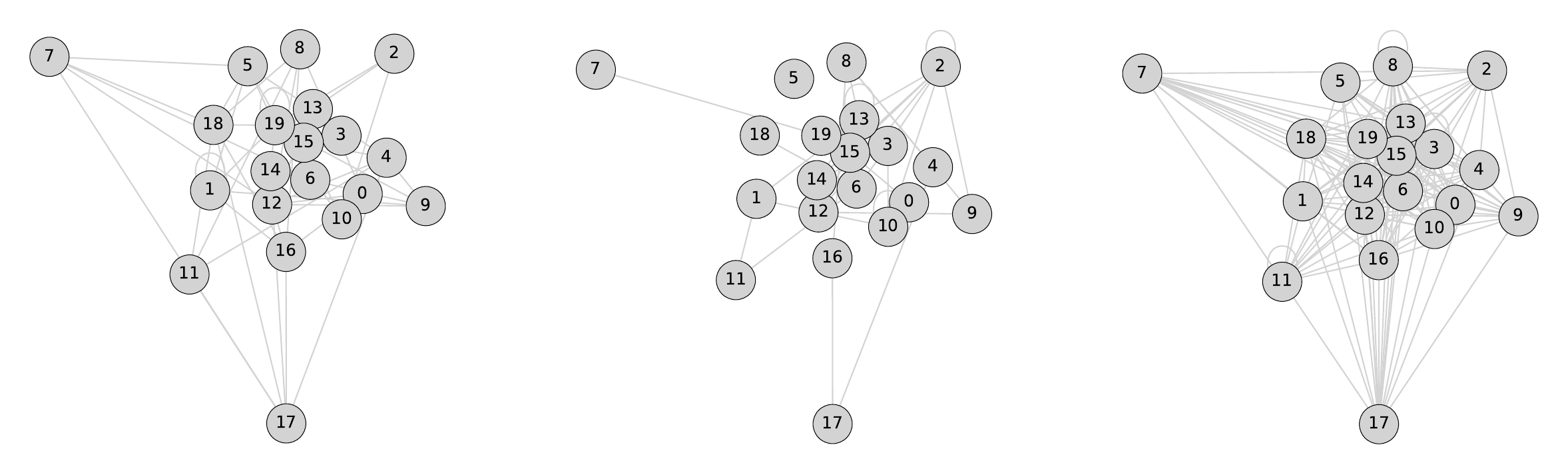}
    \caption{\textit{(Top)} The affinity matrices corresponding to the three cognitive spaces identified in the last week of the Hunter Friendship \acs{css}. \textit{(Bottom)} Three synthetic networks showing a random realization of a network generated using each of the three affinity matrices.}
    \label{fig:space_vis}
\end{figure}

In this section we highlight two possible visualizations for understanding the differences in the generative processes of each cognitive space in the last week of the Hunter Friendship \acs{css} (see \Cref{subsec:hunter}). The difference between each of the $C$ cognitive spaces is the affinity matrix describing the \acs{sbm}. As such, we first visualize the three different affinity matrices corresponding to the basis identified in considering the relative cognitive spaces. In \Cref{fig:space_vis} \textit{(top)} we visualize the affinity matrices corresponding to the perceptions of students 9, 2, and 10, respectively, each representing the basis of the three cognitive spaces. Interpreting the differences between these affinity matrices, we see that student 9's generative model assumes that the first social group has more relative in-group connections than the second social group, which is opposite of what student 2's model assumes. Student 10, conversely, assumes that the most connections occur between the first and second social group, and that all connections occur at a higher rate. In \Cref{fig:space_vis} \textit{(bottom)} we visualize one random initialization of a network generated from a \acl{sbm} with the same $\U$ and $\V$ matrices, but with affinity matrices corresponding to those in the top of the figure. Although it is difficult to interpret differences between random realizations of these models (because the randomness in an edge being present or not), we are clearly able to visualize the relatively more dense connections in Student 10's generative process.

%% file: affect_table.tex
\begin{table}
\centering
\begin{tabular}{l|l|l|l|l}
           & \textbf{A}                                                                & \textbf{B}                                                                   & \textbf{C}                                                          & \textbf{D}                                                                            \\ \hline
\textbf{A} & \begin{tabular}[c]{@{}l@{}}Opportunistic,\\ Considerate\end{tabular}      & \begin{tabular}[c]{@{}l@{}}Honest,\\ Trusting\end{tabular}                   & \begin{tabular}[c]{@{}l@{}}Strategic, \\ Profitable\end{tabular}    & \begin{tabular}[c]{@{}l@{}}Loyal, \\ Communal\end{tabular}                            \\ \hline
\textbf{B} & \begin{tabular}[c]{@{}l@{}}Self-Interested,\\ Liars/Trifling\end{tabular} & \begin{tabular}[c]{@{}l@{}}Argumentative,\\ Hop-Scotch Rational\end{tabular} & \begin{tabular}[c]{@{}l@{}}Reactive,\\ Self-Interested\end{tabular} & \begin{tabular}[c]{@{}l@{}}Altruistic,\\ Long Sighted\end{tabular}                    \\ \hline
\textbf{C} & \begin{tabular}[c]{@{}l@{}}Self-Interested,\\ Bullied\end{tabular}        & \begin{tabular}[c]{@{}l@{}}Impressionable,\\ Forgettable\end{tabular}        & \begin{tabular}[c]{@{}l@{}}Balanced,\\ Adaptive\end{tabular}        & \begin{tabular}[c]{@{}l@{}}Racheted, \\ Idealistic\end{tabular}                       \\ \hline
\textbf{D} & Filth, Judas                                                              & \begin{tabular}[c]{@{}l@{}}Conflicted,\\ Logical\end{tabular}                & \begin{tabular}[c]{@{}l@{}}Sell out,\\ Capitalist\end{tabular}      & \begin{tabular}[c]{@{}l@{}}Socially Conscious,\\ Resistant to capitalism\end{tabular}
\end{tabular}
\caption{The ``affect matrix'' from \protect\citet{hunter2019}. Each group was asked to describe the actions of the other groups following a Prisoner's Dilemma activity during the course (e.g., group B described group D as ``altruistic'' and ``long sighted.'').}
\label{tab:affect}
\end{table}

%% file: LRT_table.tex
\begin{table}
\centering
\begin{tabular}{l|l|l}
\multicolumn{1}{c|}{\textbf{Dataset}}                                                  & \multicolumn{1}{c|}{\textbf{Test}}                                                  & \multicolumn{1}{c}{\textbf{LRT Determination}}                                                  \\ \hline
\multirow{3}{*}{Krackhardt Advice}                                                     & \begin{tabular}[c]{@{}l@{}}$H_0$: Redundant \\ $H_1$: $C = 2$\end{tabular}      & \multicolumn{1}{l}{\begin{tabular}[c]{@{}l@{}}reject $H_0$\\ p-value \texttt{<1e-16}\end{tabular}}             \\ \cline{2-3}   
& \begin{tabular}[c]{@{}l@{}}$H_0$: Dependent $K=C=3$\\ $H_1$: Independent\end{tabular} & \multicolumn{1}{l}{\begin{tabular}[c]{@{}l@{}}fail to reject $H_0$\\ p-value \texttt{0.887}\end{tabular}}                                                                           \\ \cline{2-3} 
 & \begin{tabular}[c]{@{}l@{}}$H_0$: SCA $K=C=3$\\ $H_1$: Dependent $K=C=3$\end{tabular} & \multicolumn{1}{l}{\begin{tabular}[c]{@{}l@{}}reject $H_0$\\ p-value \texttt{<1e-16}\end{tabular}}             \\ \hline
\multirow{3}{*}{Krackhardt Friendship}                                                 & \begin{tabular}[c]{@{}l@{}}$H_0$: Redundant \\ $H_1$: $C = 2$\end{tabular}          & \multicolumn{1}{l}{\begin{tabular}[c]{@{}l@{}}reject $H_0$\\ p-value \texttt{<1e-16}\end{tabular}}             \\ \cline{2-3} 
& \begin{tabular}[c]{@{}l@{}}$H_0$: Dependent $K=3, C=5$\\ $H_1$: Independent\end{tabular}  & \multicolumn{1}{l}{\begin{tabular}[c]{@{}l@{}}fail to reject $H_0$\\ p-value 0.995\end{tabular}} \\ \cline{2-3} 
 & \begin{tabular}[c]{@{}l@{}}$H_0$: SCA $K=C=3$\\ $H_1$: Dependent $K=3, C=5$\end{tabular}    & \multicolumn{1}{l}{\begin{tabular}[c]{@{}l@{}}reject $H_0$\\ p-value 9.92e-13\end{tabular}}      \\ \hline
\multirow{3}{*}{\begin{tabular}[c]{@{}l@{}}Hunter Friendship,\\ Week One\end{tabular}} & \begin{tabular}[c]{@{}l@{}}$H_0$: Redundant \\ $H_1$: $C = 2$\end{tabular}          & \multicolumn{1}{l}{\begin{tabular}[c]{@{}l@{}}reject $H_0$\\ p-value \texttt{<1e-16}\end{tabular}}             \\ \cline{2-3} 
 & \begin{tabular}[c]{@{}l@{}}$H_0$: Dependent $K=2, C=3$\\ $H_1$: Independent\end{tabular}  & \multicolumn{1}{l}{\begin{tabular}[c]{@{}l@{}}reject $H_0$\\ p-value 0.004\end{tabular}}         \\ \cline{2-3} 
 & \begin{tabular}[c]{@{}l@{}}$H_0$: SCA $K=C=2$\\ $H_1$: Dependent $K=2, C=3$\end{tabular}    & \multicolumn{1}{l}{\begin{tabular}[c]{@{}l@{}}reject $H_0$\\ p-value  \texttt{<1e-16}\end{tabular}}            \\ \hline
\multirow{3}{*}{\begin{tabular}[c]{@{}l@{}}Hunter Friendship\\ Week Six\end{tabular}}  & \begin{tabular}[c]{@{}l@{}}$H_0$: Redundant \\ $H_1$: $C = 2$\end{tabular}          & \multicolumn{1}{l}{\begin{tabular}[c]{@{}l@{}}reject $H_0$\\ p-value \texttt{<1e-16}\end{tabular}}             \\ \cline{2-3} 
 & \begin{tabular}[c]{@{}l@{}}$H_0$: Dependent $K=2, C=3$\\ $H_1$: Independent\end{tabular}  & \multicolumn{1}{l}{\begin{tabular}[c]{@{}l@{}}fail to reject $H_0$\\ p-value 0.375\end{tabular}} \\ \cline{2-3} 
 & \begin{tabular}[c]{@{}l@{}}$H_0$: SCA $K=C=2$\\ $H_1$: Dependent $K=2, C=3$\end{tabular}    & \multicolumn{1}{l}{\begin{tabular}[c]{@{}l@{}}reject $H_0$\\ p-value \texttt{<1e-16}\end{tabular}}            
\end{tabular}
\caption{The p-values and \acs{lrt} determinations for the Krackhardt and Hunter CSS datasets using the standard \acs{lrt}. See \Cref{sec:krack} to compare these determinations to those of the \acs{slrt}. Of note is that the differences in the two tests are for the layer redundancy and \acs{sca} tests in the Krackhardt Friendship \acs{css}, the layer dependency test in the Hunter Week One \acs{css}, and the \acs{sca} test in the Hunter Week Six \acs{css}. In all of these tests, the \acs{slrt} failed to reject $H_0$, and the standard \acs{lrt} rejected $H_0$. This observation may be due to the \acs{slrt} being lower powered, which we discuss in \Cref{apdx:lrts}.}
\label{table:standard_LRT}
\end{table}